\documentclass[pre,twocolumn,superscriptaddress,preprintnumbers,nopacs,floatfix,amsmath,amssymb]{revtex4-1}

\usepackage{graphicx}
\usepackage{dcolumn}
\usepackage{bm}

\usepackage{amsmath}
\usepackage{graphicx}
\usepackage{amssymb}
\usepackage{mathrsfs}
\usepackage{bbm}
\usepackage{epsfig}

\newcommand{\be}{\begin{eqnarray}}
\newcommand{\ee}{\end{eqnarray}}

\begin{document}

%
%
%
\title{ Protein loops, solitons and side-chain visualization \\ with 
    applications to  the left-handed helix region 
  }

\author{Martin Lundgren}
\email{Martin.Lundgren@physics.uu.se}
\affiliation{Department of Physics and Astronomy, Uppsala University,
P.O. Box 803, S-75108, Uppsala, Sweden}
\author{Antti J. Niemi}
\email{Antti.Niemi@physics.uu.se}
\affiliation{
Laboratoire de Mathematiques et Physique Theorique
CNRS UMR 6083, F\'ed\'eration Denis Poisson, Universit\'e de Tours,
Parc de Grandmont, F37200, Tours, France}
\affiliation{Department of Physics and Astronomy, Uppsala University,
P.O. Box 803, S-75108, Uppsala, Sweden}
\author{Fan Sha}
\email{fansha0559@gmail.com}
\affiliation{Department of Physics and Astronomy, Uppsala University,
P.O. Box 803, S-75108, Uppsala, Sweden}

\begin{abstract}
\noindent
Folded proteins have a modular assembly. They are constructed from regular
secondary structures  like $\alpha$-helices 
and $\beta$-strands that are  joined together by  loops.   
Here we develop a visualization technique that is adapted to describe
this  modular structure.  In complement to the widely employed 
Ramachandran plot that is based on toroidal geometry, our approach 
utilizes the geometry of a two-sphere. Unlike the more conventional approaches that only describe
a given peptide unit, ours is capable of describing the entire backbone environment  including the
neighboring peptide units. It
maps the positions of  each atom  to the surface of the two-sphere exactly    
how these atoms  are seen by an observer who is located at the position of the central $C_\alpha$ atom.
At each level of side-chain  atoms
we observe a strong  correlation between the positioning of the atom and the underlying 
local secondary structure with very little if any variation between the
different amino acids. As a concrete example we analyze  the left-handed helix region of  
non-glycyl amino acids.  
This region  corresponds to an isolated and highly localized  residue independent  sector in the direction of 
the  $C_\beta$ carbons on the two-sphere.  We show that the residue independent  localization extends 
to   $C_\gamma$ and $C_\delta$ carbons,  and to side-chain oxygen and nitrogen atoms 
in the case of asparagine and aspartic acid. When we extend the analysis 
to the side-chain atoms of the neighboring residues, we observe that left-handed $\beta$-turns 
display a regular and largely amino acid independent  structure that can extend  to  seven 
consecutive residues.  This collective  pattern is due to the presence of a backbone soliton. 
We show how one can use our visualization techniques to analyze and classify the different 
solitons in terms of selection rules that we describe in detail.
\end{abstract}



\maketitle

\section{Introduction}
The Ramachandran plot \cite{rama1}, \cite{rama2} is the paradigm technique of protein visualization.
It describes backbone atoms in a peptide group around a given C$_\alpha$ carbon 
in terms of dihedral rotations. Ramachandran
plot can also been extended to the side-chain atoms in terms of  the dihedral rotamers. This
gives rise to Janin plot \cite{janin} and its variants. 
In the present article we develop new visualization techniques to describe proteins. Our goal is
to visualize all atoms both in a given peptide unit and those in the neighboring units, beyond the regime
of the Ramachandran plot. This will
enable us to  search for new  relations between the positioning 
of various atoms and the backbone geometry. Our approach draws from developments in three 
dimensional visualization that have taken place after the Ramachandran  plot was originally
introduced \cite{hansonbook}, \cite{kuipers}. In particular, in lieu of toroidal geometry  
we utilize the geometry of a two-sphere. It enables us to describe the 
various atoms exactly as they are seen by an observer who roller-coasts along the backbone. 
Of particular interest to us is the visual analysis of the modular components of which all
folded proteins are built.  These have been recently  identified as the soliton solutions to a generalized
discrete nonlinear Schr\"odinger equation (DNLS) \cite{cherno}-\cite{peng}.

Soliton solutions to nonlinear difference equations 
share a long history with biological physics of proteins.
The discrete version of the nonlinear Schr\"odinger equation is an embodiment of this  
relationship. It 
was originally introduced by Davydov \cite{davy} to describe  energy transfer along the protein
$\alpha$-helices. Subsequently the DNLS equation has found many additional applications  
in biological physics and  elsewhere \cite{kevk}.  The DNLS equation has also the remarkable mathematical
property of integrability,  it is commonly viewed as the archetype integrable system \cite{fadd}. 

When the DNLS soliton propagates along the $\alpha$-helix,  
the protein changes its shape. In \cite{cherno},  \cite{nora} it has been shown that when
the soliton becomes trapped,  the protein folds.  It now appears that practically all 
folded proteins can be built in a modular fashion from a relatively small number of such trapped  solitons 
\cite{peng}.  In the present article we combine the notion of  soliton with modern visualization techniques 
\cite{dff}.  We are particularly interested in the ramifications  of the backbone DNLS soliton  in
protein side-chain geometry.  Our  ultimate goal is  to develop a graphical characterization  and eventually a full classification 
of protein structures in terms of their soliton modules.  As a prelude, we here utilize the soliton concept  
to visually inspect and analyze those protein conformations that are
located in the  left handed $\alpha$-helix ($\tt L$-$\alpha$) region of the Ramachandran plot \cite{rama1}, \cite{rama2}. 
This region is a relatively small subset of all different protein conformations, and as such amenable to an
explicit analysis.

Of particular interest to us are  the common  geometric aspects of the asparagine (ASN) and  aspartic acid 
(ASP).  Asparagine  is the predominant residue in the so-called non-glycyl  
$\tt L$-$\alpha$ region.    According to  the prevailing point of  
view this is due to a {\it localized}  non-covalent attractive carbonyl-carbonyl interaction between the 
side-chain and  backbone  \cite{deane}-\cite{review}.  Such a carbonyl-carbonyl interaction can only be present
in ASN, ASP, glutamine (GLN) and glutamic acid (GLU).  Indeed, the propensity of  ASP  that is structurally very
similar to ASN is also clearly amplified in the $\tt L$-$\alpha$ region, while the somewhat lower propensity of GLN and GLU  
has been explained in the literature  to be a consequence of steric suppressions \cite{deane}.  

Here we show that the presence of a $\tt L$-$\alpha$ site goes beyond the regime of a single peptide unit. We find
that it involves a coordinated 
interplay of up to seven  consecutive amino acids. We argue that this extended correlation over several amino acids is 
symptomatic to solitons. We perform a detailed visual investigation and propose a graphical classification
of these  solitons. We argue that all protein structures could be characterized and classified similarly,
in terms of  general  selection rules that we formulate. We find that the continuous 
geometry of the two-sphere  gives a more perceptible characterization of protein 
conformations than the toroidal Ramachandran plot. In fact,  the  three dimensional 
visualization techniques  we  utilize have been largely introduced and  developed after the 
publication of \cite{rama1}, \cite{rama2}.  Our  approach exploits  the properties  of a  piecewise 
linear {\it framed }  chain, as it is being applied to visualization problems 
in aircraft and robot kinematics,  stereo reconstruction, and increasingly in
computer graphics and virtual reality \cite{hansonbook}, \cite{kuipers}. 
In these applications different framings  correspond to different camera gaze positions, that one introduces  and varies
for the purpose of extracting diverse and complementary information on geometrical
aspects and physical properties  of the system under investigation. However, largely due to the 
success and systematics provided by Ramachandran plot, thus  far this kind of  approach 
has been sparsely applied to the analysis of protein conformations.  Among our goals is to demonstrate  
that these modern visualization techniques can provide  a powerful complementary  tool 
for  the  visual description of folded proteins.  In particular, they enable the study of visual 
correlations between nearby peptide units, which is not possible in the Ramachandran approach that is limited
to to describe a single peptide unit only.

Finally, we note that the  investigation of the physical properties  of our concrete examples ASN and ASP is also of  substantial 
biological  interest. These two amino acids are more frequently than any other amino acid subject 
to {\it in vivo} post-translational 
modifications including spontaneous nonenzymatic deamidation from ASN to ASP \cite{deami} 
and racemization from  $\tt L$-ASP into $\tt D$-ASP \cite{race}. These processes are presumed to have consequences  
to  cellular and organismal  ageing \cite{deami}, \cite{deami2}. They  might also have a r\^ole in  enhancing the emergence of amyloid based neurodegenerative diseases \cite{deami2},  \cite{prion}.



\section{Framing}

We interpret a protein backbone  in terms of framed chain, with vertices located at the $C_\alpha$ carbons \cite{dff}. 
Depending on the application, the framing can  be introduced in various different ways. Examples include the geometric Frenet frame \cite{hansonbook}, \cite{kuipers},  the geodesic Bishop frame \cite{bishop}, and
protein specific $C_\beta$ carbon frame that we obtain by utilizing the direction of the C$_\beta$ carbon along a protein backbone to construct an orthonormal framing \cite{dff}.  Here we propose that  in particular the Frenet framing provides a powerful tool for protein side-chain visualization, also beyond our 
explicit example of the $\tt L$-$\alpha$ Ramachandran region. The additional advantage of the Frenet framing is that it
relates directly to an energy function. But we also advertise the closely related  C$_\beta$
framing that may sometimes have certain visual advantages. 

The framing of a piecewise linear chain is conventionally  based
on the Denavit-Hartenberg \cite{dh} formalism. This formalism  was  originally introduced  in robotics but has been subsequently
extensively applied also in other disciplines. 
Here we resort to a variant,  that has been developed in \cite{dff} for the purpose of framing protein backbones.  It  
utilizes the transfer matrix formalism  \cite{fadd} to
describe a protein with $N$ residues using the coordinates $\mathbf r_i$ of the backbone $C_\alpha$ 
carbons  ($i=1,...,N$). These coordinates can be downloaded from 
the Protein Data Bank (PDB) \cite{pdb}.
For each of the segments that connect the backbone $C_\alpha$ central  carbons  we compute  the unit 
length tangent vector $\mathbf t_i$, binormal vector $\mathbf b_i$ and normal vector $\mathbf n_i$ using
\[
{\bf t}_i = \frac{ {\bf r}_{i+1} - {\bf r}_i }{ | {\bf r}_{i+1} - {\bf r}_i |}
\]
\begin{equation}
{\bf b}_i = \frac{ {\bf t}_{i-1} \times {\bf t}_i }{| {\bf t}_{i-1} \times {\bf t}_i|} 
\label{tbn}
\end{equation}
\[
{\bf n}_i = {\bf b}_i \times {\bf t}_i 
\]
Thus the tangent vector $\mathbf t_i$ points from the $i^{th}$ central $C_\alpha$ carbon 
 to the direction of the $(i+1)^{th}$ central $C_\alpha$ carbon, the way how it is seen by an observer  who is located at  the position of
the $i^{th}$ carbon. The $\mathbf b_i$ and $\mathbf n_i$ determine a  frame that enables the observer  to orient herself 
at the location $\mathbf r_i$, on the
plane that is orthogonal to the direction $\mathbf t_i$. Together the right-handed  triplet $(\mathbf n_i , \mathbf b_i , \mathbf t_i)$ 
constitutes the orthonormal 
discrete Frenet frame  for each residue along the backbone chain, with base  at the position of the vertex $\mathbf r_i$.  
The corresponding backbone bond $\kappa_{i+1,i} \equiv \kappa_i$ and torsion $\tau_{i+1,i} \equiv
\tau_i$ angles can  be computed from (\ref{tbn}) as follows,
\begin{equation}
\cos \kappa_i \ = \ {\bf t}_{i+1} \cdot {\bf t}_i
\label{kappa}
\end{equation}
\begin{equation}
\cos \tau_i \ = \    {\bf b}_{i+1} \cdot {\bf b}_i
\label{tau}
\end{equation}
Alternatively, if the bond and torsion angles are known we can construct the frames iteratively by starting from the $N$ terminus and using  \cite{dff}
\[
\left( 
\begin{matrix} 
{\bf n}  \\  {\bf b } \\ {\bf t}
\end{matrix} \right)_{i+1} \! \! \!
= \  {\mathcal R}_{i+1,i} 
\left( \begin{matrix} {\bf n} \\  {\bf b } \\ {\bf t} \end{matrix} \right)_i 
\]
\begin{equation}
= \  \exp\{ \kappa_{i} T^2  \}
\cdot \exp \{ \tau_{i} T^3 \} \! \left( \begin{matrix} {\bf n} \\  {\bf b } \\ {\bf t} \end{matrix} \right)_i 
\label{DFE}
\end{equation}
where the $T^a$ $(a=1,2,3)$  are the adjoint SO(3) Lie algebra generators. Once the $\mathbf t_i$ have been
constructed from (\ref{DFE})  and the
bond lengths $s_i = |{\bf r}_{i+1} - {\bf r}_i|$ have been determined  we recover
the entire backbone  from
\begin{equation}
{\bf r}_k \  = \ \sum_{i=0}^{k-1} s_i \cdot {\bf t}_i 
\label{back}
\end{equation}
The set of all Frenet frames  defines a framing of the backbone.
According to (\ref{kappa})-(\ref{DFE}) the bond and torsion angles are link variables, they relate a frame at the vertex $\mathbf r_i$ to a frame
at the vertex $\mathbf r_{i+1}$. We note that the definition of the bond angle 
involves three vertices while the definition of the torsion angle involves a total of four vertices. 

The C$_\beta$ framing \cite{dff} is a complement to the Frenet framing. It can be introduced
for all non-glycyl residues. We define 
these frames  similarly, in terms of three mutually  orthogonal unit vectors at each C$_\alpha$ carbon. 
Consequently the Frenet framing and the C$_\beta$ framing are related to each other by peptide unit 
dependent SO(3) rotations. The first unit vector of the C$_\beta$ basis is obtained as follows,
\[
\mathbf s_i  \ = \ \frac{ \mathbf r_{\beta,i} - \mathbf r_{\alpha,i} }{
|\mathbf r_{\beta,i} - \mathbf r_{\alpha,i}|}
\]
Here $\mathbf r_{\alpha,i}$ is the location of the $i$th C$_\alpha$ atom, and  $\mathbf r_{\beta,i}$
is the location of the corresponding C$_\beta$ atom. The second unit vector is
\[
\mathbf p_i \ = \ \frac{ \mathbf s_i \times \mathbf t_i }{ |\mathbf s_i \times \mathbf t_i |}
\]
where $\mathbf t_i$ is the Frenet frame unit tangent vector. Finally, the third unit vector in the
C$_\beta$ frame is
\[
\mathbf q_i \ = \ \mathbf s_i \times \mathbf p_i
\]
Since ($\mathbf s_i, \mathbf p_i, \mathbf q_i$) is an orthonormal frame located at each C$_\alpha$,
it can be used like the Frenet frame to visualize the various atoms along the protein backbone.
Moreover, since 
\[
\mathbf t_i = \mathbf p_i \times \mathbf s_i 
\]
we can likewise use the C$_\beta$ framing to construct the entire backbone using (\ref{back}).

We wish to employ the various frames together with the 
discrete Frenet equation (\ref{DFE}) to inspect  the structure of folded  proteins.  As 
our principal data set 
we utilize all those proteins that are presently in PDB and have an overall resolution that is better
than 2.0 \.Angstr\"om. We introduce no additional curation or data pruning in this set.  But we have 
confirmed that all our results and conclusions stand when we restrict ourselves to
those proteins with resolution better than 1.5 \.A and with less than 30$\%$
homology relation, or to those proteins that have a resolution which is better than 1.0 \.A. Finally, as 
a control set  we also utilize the highly curated version v3.3 Library of chopped PDB files for representative 
 CATH domains \cite{cath}. Since the conclusions we  draw are indifferent of the data set that we use  
 we only describe explicitly  the results for the  first one, as it allows for the visually most complete presentation.



\section{Backbone mapping}

We start by describing how to visualize the  protein backbone in terms of the Frenet frames \cite{dff}. 
Here we go beyond the regime of 
the Ramachandran plot, that does not provide any direct visual correlation between neighboring peptide groups.
We introduce an observer who maps all  the atoms in the protein 
by traversing along the backbone. The observer  
moves between the $C_\alpha$ carbons like on a roller-coaster  with an orientation that is determined
by the discrete Frenet framing:
We take the base of the tangent vector $\mathbf t_i$ defined  in (\ref{tbn}) to be at the location $\mathbf r_{i}$  of the $i^{th}$ 
central  $C_\alpha$ carbon.  The tip of $\mathbf t_i$  then determines a point on the surface  of a unit two-sphere that surrounds our observer at the location
of this $C_\alpha$ carbon. The observer uses this two-sphere to  
constructs a map of the various atoms exactly the way 
how she sees them on the surface of the sphere, as if the atoms were stars in the sky. 
For this she  always orients the two-sphere  at the site $i$ so that the north-pole coincides with the
tip of $\mathbf t_i$ {\it i.e.}  the north-pole is always in the direction of the next $C_\alpha$ at the site $\mathbf r_{i+1}$. She
takes the bond  angle to measure the latitude of the two-sphere from its north pole. The torsion angle  measures  the
longitude starting from the great circle that passes both through the north pole and through the tip of the binormal vector $\mathbf b_i$.
In terms of these angles she can characterize  the direction  of the vector $\mathbf t_{i+1}$ {\it i.e.} the 
direction towards site $\mathbf r_{i+2}$ to which the roller coaster turns at the next $C_\alpha$ carbon. Consequently
she acquires information about the geometric relations between neighboring peptide units, 
and this goes beyond the regime of the Ramachandran plot. She proceeds as follows:  

She first translates the center of the  two-sphere from the location of the $i^{th}$ central carbon towards its north-pole and all the way to the location of the $(i+1)^{th}$ central 
carbon, without introducing any rotation of the sphere. She  then records the direction of $\mathbf t_{i+1}$  
as a point on the surface of the  two-sphere. This defines the corresponding coordinates ($\kappa_i, \tau_i$)  and marks
a point on the map.  It gives an instruction to the observer at the point $\mathbf r_i$,  how she should turn
at site $\mathbf r_{i+1}$, to reach  the  $(i+2)^{th}$ central $C_\alpha$ carbon at the point $\mathbf r_{i+2}$. 

She then continues to construct  the mapping with the next $C_\alpha$ carbon along
the backbone. She rotates the two-sphere at $\mathbf r_{i+1}$ so that the north pole of the rotated sphere coincides with the tip of  $\mathbf t_{i+1}$, and so that the torsion angle measures the longitude from the great circle determined by the north-pole and the tip of  
$\mathbf b_{i+1}$.  She repeat the procedure for all $C_\alpha$, until she has mapped  the entire backbone. 
We note that for a folded protein the two vectors $\mathbf t_i$ and $\mathbf t_{i+1}$ are never exactly parallel to each other so there is never any ambiguity due to an inflection point. 

When we repeat this mapping procedure for every $C_\alpha$ in all proteins in our data set,  we 
obtain a ($\kappa, \tau$) distribution that characterizes the overall geometry of protein backbones.  
This provides non-local information on the backbone geometry that extends over several peptide units.
In particular, we now have a map that
shows {\it exactly}  how the central carbons are seen by our
roller-coasting observer when she gazes at them from her Frenet frame
positions along the backbone.   

We find that the $C_\alpha$ distribution for all proteins in our data set determines an annulus on the surface of 
the two-sphere. For visualization it then becomes convenient to employ  the geometry of the stereographically 
projected  two-sphere. It  is obtained by projecting our $(\kappa, \tau)$ coordinates to the north pole tangent plane of the two-sphere. If ($x,y$) are the coordinates of this tangent plane the projection is defined by 
 \begin{equation}
 x + iy =  \tan  (\frac{\kappa}{2}) \cdot e^{- i \tau}
 \label{stere}
 \end{equation}
 When we perform this projection for all $C_\alpha$ carbons in all proteins that are 
 in our data set and separately display the results for the different groups of $\alpha$-helices, 
 $\beta$-strands, $3/10$-helices and loops as these structures are defined in PDB, 
 we arrive at the angular  distributions that we show  in Figures 1.   
 For our observer who always fixes her gaze position towards the north-pole 
 of the surrounding two-sphere at each $C_\alpha$ carbon,  {\it i.e.}  towards the black dot at the center of the annulus,  the color intensity reveals
 the likely direction to which the roller coaster who is located at position $\mathbf r_i$ turns at the next $C_\alpha$ carbon, when she starts moving from its location at
 $\mathbf r_{i+1}$ towards $\mathbf r_{i+2}$.  In particular, the four maps in Figure 1 are  
 in a direct visual correspondence with
the way how the Frenet frame observer perceives the backbone geometry.  


 \begin{figure}[!hbtp]
  \begin{center}
    \resizebox{7.5cm}{!}{\includegraphics[]{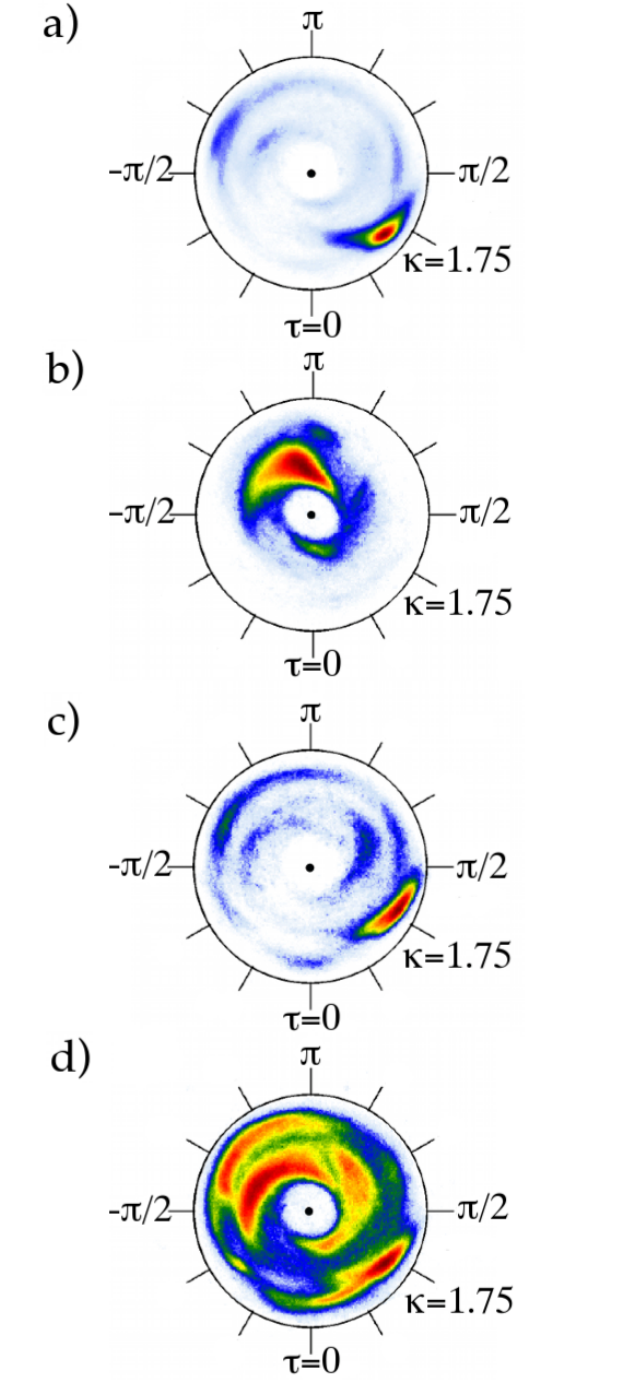}}
    \caption{(Color online:) The four major protein structures: a) 
$\alpha$-helices,  b) $\beta$-strands,   c) $3/10$-helices and d) loops. We define these structures  according to their
PDB classification in our 2.0 \.A data set.
In each Figure the center of the annulus is  the north-pole of the two-sphere  that surrounds the observer at the position $i$.
This is the direction where the next $C_\alpha$ is located. 
At this point the bond (latitude) angle $\kappa = 0$. 
The bond angle measures distance from the center of the annulus so that 
the south pole  where $\kappa=\pi$ corresponds to points at infinity on the plane.  The torsion {\it i.e.}  longitude 
angle $\tau\in [-\pi,\pi]$  increases by $2\pi$ when we go around the center of the annulus in counter-clockwise direction. 
The color coding in all our Figures  increases from white to blue to green to yellow to red  describes the relative number of conformations in 
PDB in a $\log$-squared scale.  The intensity is proportional to the probability of  the 
direction where the observer turns at the next $C_\alpha$ carbon.} 
    \label{fig:figure-1}
  \end{center}
\end{figure}


The four maps in Figure 1 portray non-local features that are not available in conventional Ramachandran plots. 
Moreover, instead of a discontinuous toroidal 
square as in the case of the Ramachandran plots, the predominant feature in all of the present maps is  that the 
PDB data is concentrated in a continuous annulus which is roughly between the circles $ \kappa_{in} \approx 1$ and $\kappa_{out}
\approx  \pi/2$. 
The exterior of the 
annulus $\kappa > \kappa_{out}$ is an excluded region, it describes  conformations that are subject to steric clashes. The  
interior $\kappa <  \kappa_{in}$ is  sterically allowed but practically excluded as 
long as proteins remain in the collapsed phase; The interior region becomes occupied when we cross  the  $\Theta$-point 
and proteins  assume their unfolded conformations.  

We notice that  loops 
appear to have a slightly higher tendency to bend towards left  {\it i.e.}  $\tau < 0$.
We also note that in the Figures for  $\alpha$, $\beta$ and $3/10$  the blue regions correspond to residues where
the present hydrogen-bond based PDB classification  is in a miss-match with the  geometric structure that is commonly associated with these configurations.
Moreover, the Figure  reveals that the PDB data displays innuendos of various 
underlying reflection symmetries: In the  Figure 1d (loops) there is a clearly visible 
mirror of the standard right-handed $\alpha$-helix region, located in the vicinity of the outer 
rim with $\kappa \approx 3/2 $ and with torsion angle close to the value $\tau \approx -2\pi/3$. A helix in this regime would be 
left-handed and tighter than the standard $\alpha$-helix. There is also
a clear mirror structure in the Figure 1b for $\beta$ strands,  the standard region is $(\kappa, \tau) \approx ( 1,\pi)$  and its less populated 
mirror is located around
$(\kappa, \tau) \approx (1,0)$. The mirror symmetry between the ensuing extended regions persists in the Figure 1d for loops. 
Finally, in the Figure 1d
we observe  a small elevated (yellow)
region in the vicinity of $(\kappa, \tau) \approx (3/2 , -\pi/3)$. 
This is the region of helices that are spatial left-handed mirror images of the standard $\alpha$-helices. There is also 
a (slightly) elevated (green) mirror of this region around $(\kappa, \tau) \approx (3/2, 2\pi/3)$.  This is like 
the  $(\kappa, \tau)  \approx (3/2 \, ,  -2\pi/3)$ mirror of the 
standard right-handed $\alpha$ helices.



\section{Side-Chain mapping}

We can similarly visualize the  geometry of side-chain atoms, as  they are seen by our roller-coasting observer. 
This gives us local information on the given peptide unit. Now the results turn out to be isomorphic to those revealed by the
standard Ramachandran plot. Moreover, this enables us to develop a visual complement to the existing rotamer libraries.

We assume that the observer is oriented according to 
the discrete Frenet framing that is determined by the transfer matrix (\ref{DFE}) at each $C_\alpha$. 
At the location of the $C_\alpha$   the 
observer then looks  at the side-chain atoms and records the direction of each of them as points on the surface of the 
two-sphere that surrounds the  observer, 
with the north-pole of the sphere always coinciding with the direction towards
the next $C_\alpha$ exactly as in the case of the backbone.

In Figure 2 (top) we display the angular distribution  of the $C_\beta$ carbons on the surface of the two-sphere
for all  the $C_\alpha$ carbons, as recorded by our Frenet frame observer who is located at the origin of the sphere.
Recall that a $C_\beta$ carbon is present in all non-glycyl residues.
We note that our framing is determined entirely in terms of the backbone. According to  prevailing paradigm the
directions of the $C_\beta$ carbons should then be  directly computable from the geometry of the tetrahedral 
covalent bond  structure of the pertinent $C_\alpha$ carbon.   However,  Figure 2 (top) reveals that the directions of 
the $C_\beta$ carbons are not determined only by the local covalent bond structure. In addition, these directions are clearly 
subject to secondary structure dependent but amino acid independent nutations.  
This confirms  that at the level of accuracy of our data, the stereochemical restraints fail to be  fully universal.
They reflect the secondary structure environment \cite{sch}-\cite{mart}. In fact, despite being based entirely
on the C$_\beta$ atoms the Figure 2 is fully isomorphic to
the standard Ramachandran plot, for all amino acids except for glycine that has no C$_\beta$.
\begin{figure}[!hbtp]
  \begin{center}
    \resizebox{8.cm}{!}{\includegraphics[]{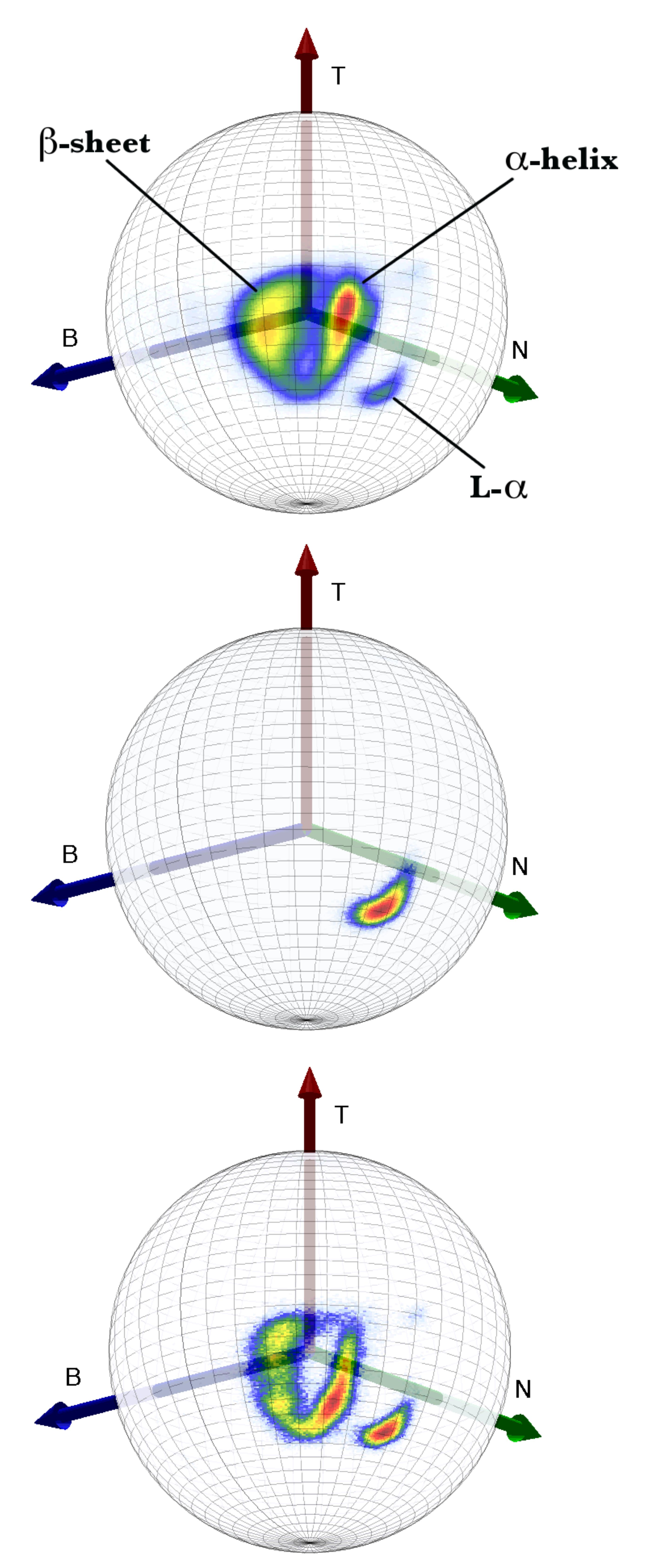}}
    \caption{(Color online:) The directions of the $C_\beta$ carbons, as seen by our Frenet frame observer who is located at  the corresponding $C_\alpha$
    carbon which is at the center of the sphere. The vector $\mathbf t$ points to the direction of the next $C_\alpha$ carbon.
    On top all residues in our data set including ASN. In the middle we display only the $\tt L$-$\alpha$
    region of the Ramachandran plot. On bottom
    we display  only those ASN  that are in a loop in PDB classification.   }
    \label{fig:figure-2}
  \end{center}
\end{figure}

A important feature of the nutation is 
the presence of the  highly localized, isolated  island denoted  $\tt L$-$\alpha$ that is clearly visible in Figure 2 (top).
We have confirmed that   this  isolated island coincides  exactly with the  conventional
non-glycyl $\tt L$-$\alpha$ region of the standard Ramachandran plot. 
This is shown in Figure 2 (middle) where we display the direction of the $C_\beta$ carbons solely  for those non-glycyl residues that are in the $\tt L$-$\alpha$  Ramachandran region.
Finally, in the Figure 2 (bottom) we display the discrete Frenet frame distribution of the $C_\beta$ carbons for those ASN that are located
in loops only, according to PDB classification.  The relatively
high propensity of ASN  in the $\tt L$-$\alpha$ island  is prominent.

In the sequel we shall concentrate our attention solely on the isolated  
$\tt L$-$\alpha$ island in Figure 2 (middle). 
We start by noting the propensity of different amino acids  in the $\tt L$-$\alpha$ island.  
The result (in percent) is shown in Figure 3. 
This Figure confirms the high propensity of ASN (N) that is also visible from Figure 2 (bottom).  We find that ASP (D) has also 
relatively high relative propensity. But  the propensity of  histidine (H) is practically  equal. Furthermore, several non-carbonylic
amino acids have a higher propensity than GLU (E).   
Finally, the  $\beta$-branched isoleucine (I), valine (V) and threorine (T)  all have
clearly suppressed propensities and proline (P) is practically absent, presumably reflecting 
the presence of steric constraints \cite{deane}, 
\cite{review}. 


 \begin{figure}[!hbtp]
  \begin{center}
    \resizebox{8.3cm}{!}{\includegraphics[]{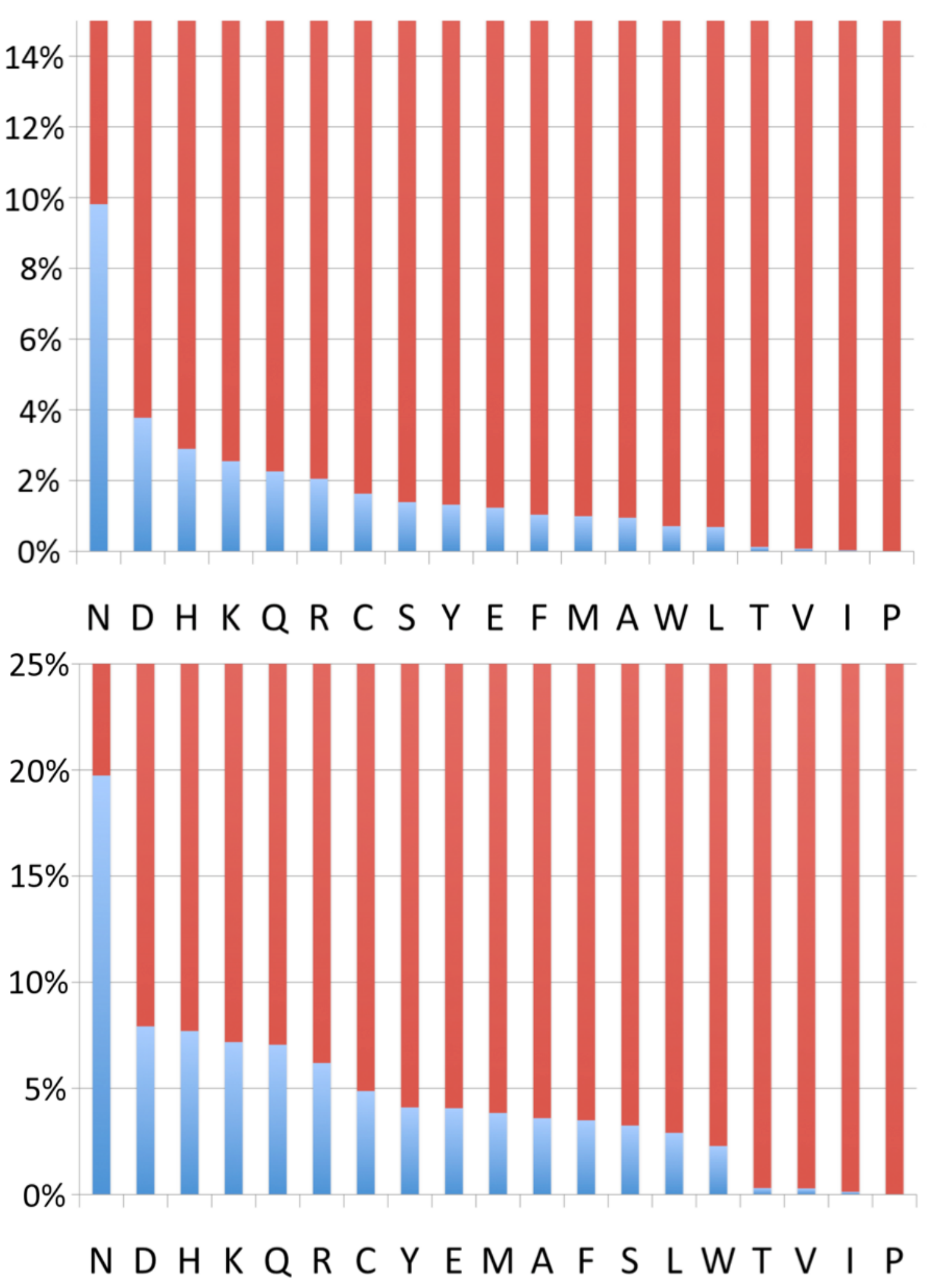}}
    \caption{(Color online:) The percent distribution of non-glycyl  residues in the $\tt L$-$\alpha$ island of Figure 2. In the top
    Figure we display the result for
    all amino acids in our entire data set, and in the bottom Figure  
    for those in our data set that are classified as loops in PDB. The propensity of
     carbonylic ASN (N) is clearly enhanced in both cases.  But in both cases the similarly carbonylic ASP (D) 
     has about the same percent-wise propensity 
     with the non-carbonylic HIS (H), and the  carbonylic GLU (E) is relatively quite suppressed. 
       }
    \label{fig:figure-3}
  \end{center}
\end{figure}


We now proceed to  map the directions of the $C_\gamma$ carbons for those side-chains where $C_\beta$ is located in 
the $\tt L$-$\alpha$ island of Figure 2.  We continue to utilize the framing determined by 
our  observer who roller-coasts the $C_\alpha$ 
backbone with orientation determined by the discrete Frenet frames, and north-pole always in the direction of the next
C$_\alpha$.  The  result is presented in Figure 4. 
It reveals 
that at the level of $C_\gamma$,   the single $\tt L$-$\alpha$ island of the $C_\beta$ becomes divided into two 
separate but still  {\it highly} localized islands. This reflects the sp3 hybridization of the C$_\beta$:
There is a putative  {\it gauche}-  ($g$-) island
where around 70$\%$ of the residues  in the $\tt L$-$\alpha$ island are located, and a putative  {\it trans} 
island for the rest.  Interestingly, we do not really see any  putative  {\it gauche}$+$ ($g+$)  island. 

The amino acid  propensities of these two islands is displayed in Figure 5.
ASN is the most populous  in both $C_\gamma$ islands. However, the propensity of ASP is elevated
only in  {\it trans} island. In the  {\it g}- island both non-carbonylic HIS  (H) and LYS (K) and even the carbonylic GLN (Q) have a 
higher propensity than ASP.  At the moment we  have no good explanation for this observation, and we leave
it as challenge.
\begin{figure}[!hbtp]
  \begin{center}
    \resizebox{8cm}{!}{\includegraphics[]{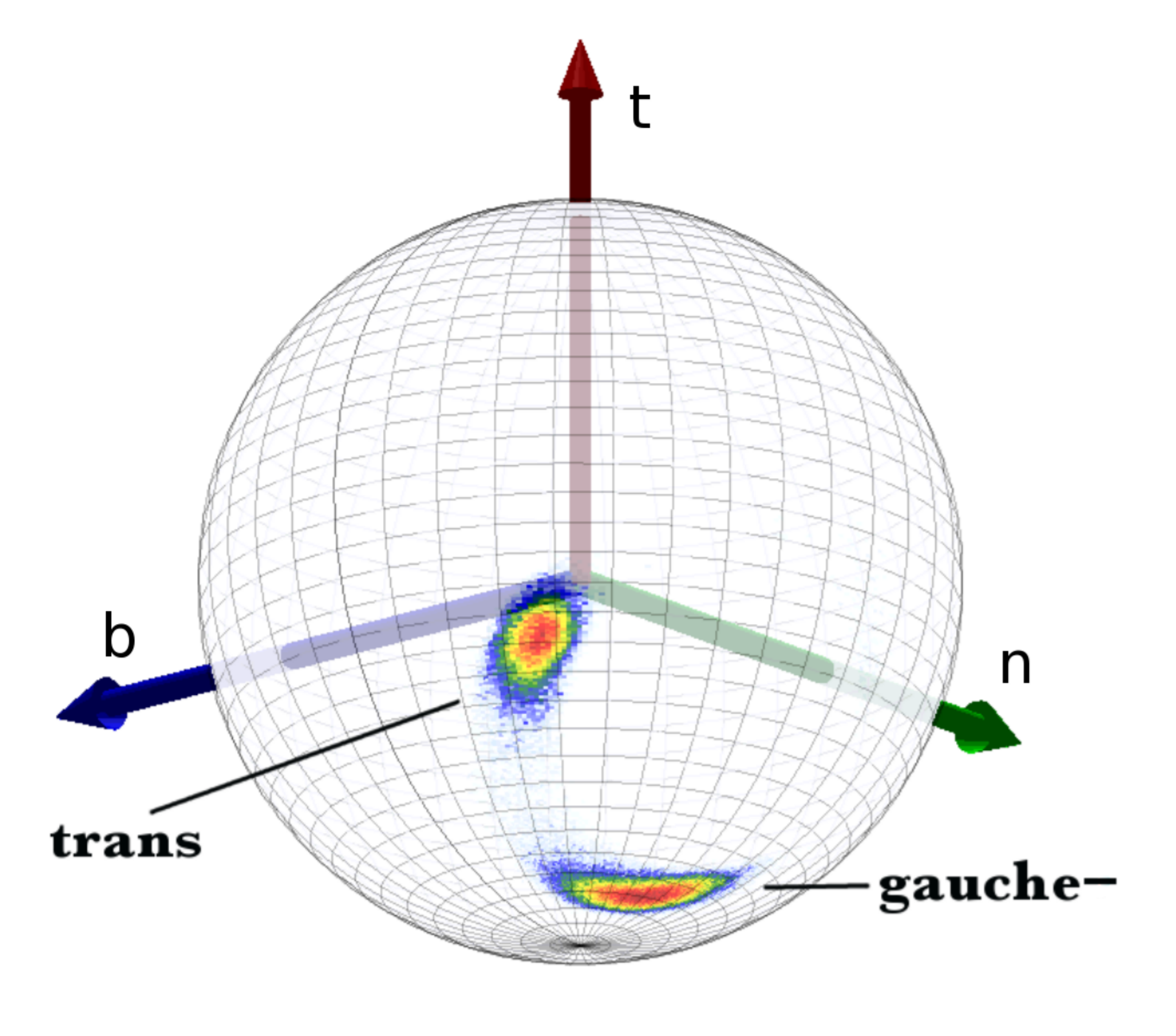}}
    \caption{(Color online:) The directions of those $C_\gamma$  carbons for which the 
    $C_\beta$ is located  in the  $\tt L$-$\alpha$ island, and as seen
    by our Frenet frame observer who is located at the $C_\alpha$ carbon which is situated at the center of the sphere.  
       }
    \label{fig:figure-4}
  \end{center}
\end{figure}
\begin{figure}[!hbtp]
  \begin{center}
    \resizebox{8cm}{!}{\includegraphics[]{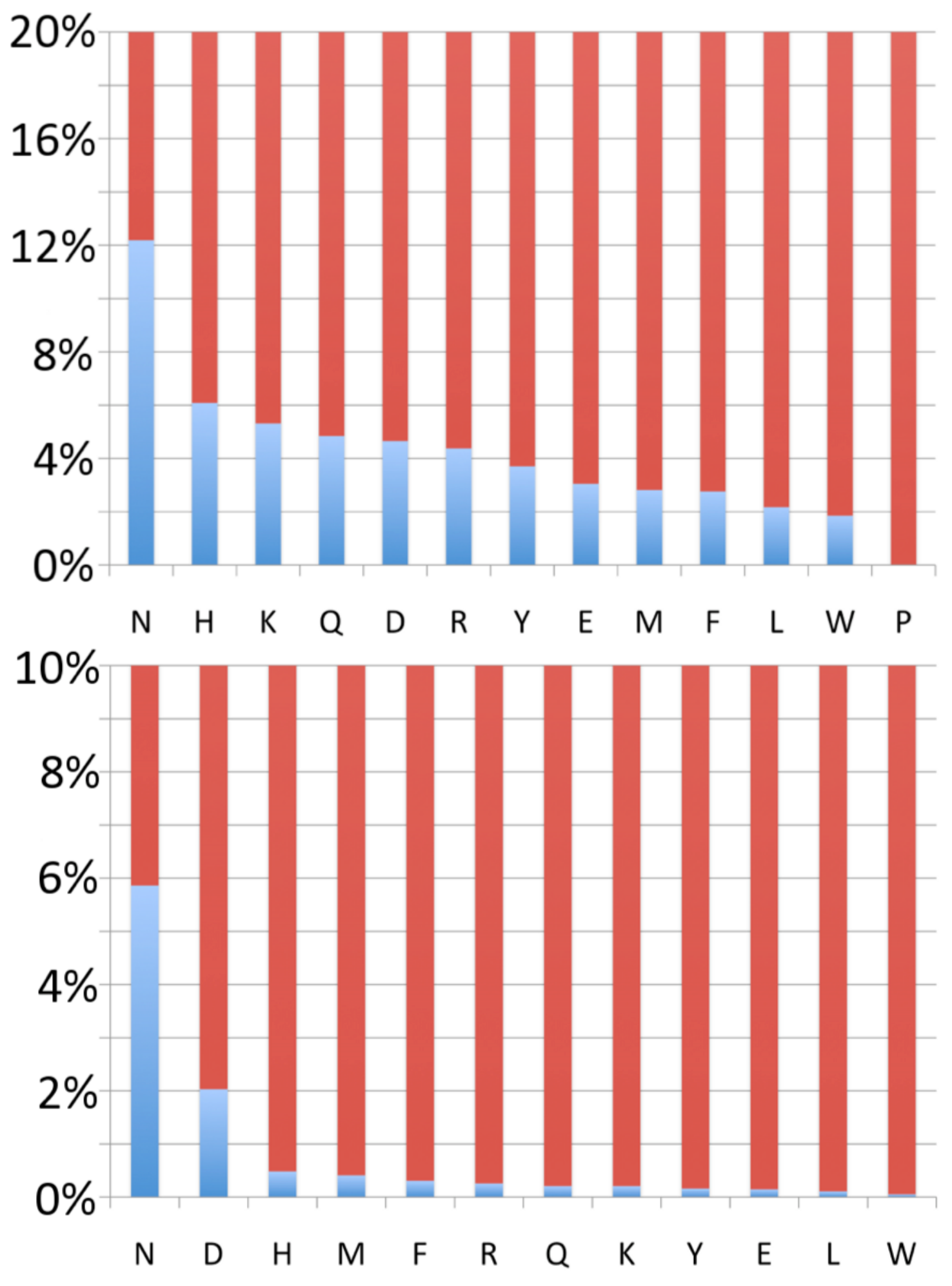}}
    \caption{(Color online:) The percent-wise  propensity of different amino acids in the putative {\it g}- island (top) and {\it trans} $C_\gamma$-island (bottom) in Figure 4}
    \label{fig:figure-5}
  \end{center}
\end{figure}
\noindent 
In Figure 6 we plot the percentage ratios of the different amino acids as they appear in the two 
$C_\gamma$ islands. We note that around $43\%$ of residues  in the putative $\it g-$ island  are 
non-carbonylic, while in the putative $\it trans$ island the number is much lower, close to $12\%$. 


\begin{figure}[!hbtp]
  \begin{center}
    \resizebox{8cm}{!}{\includegraphics[]{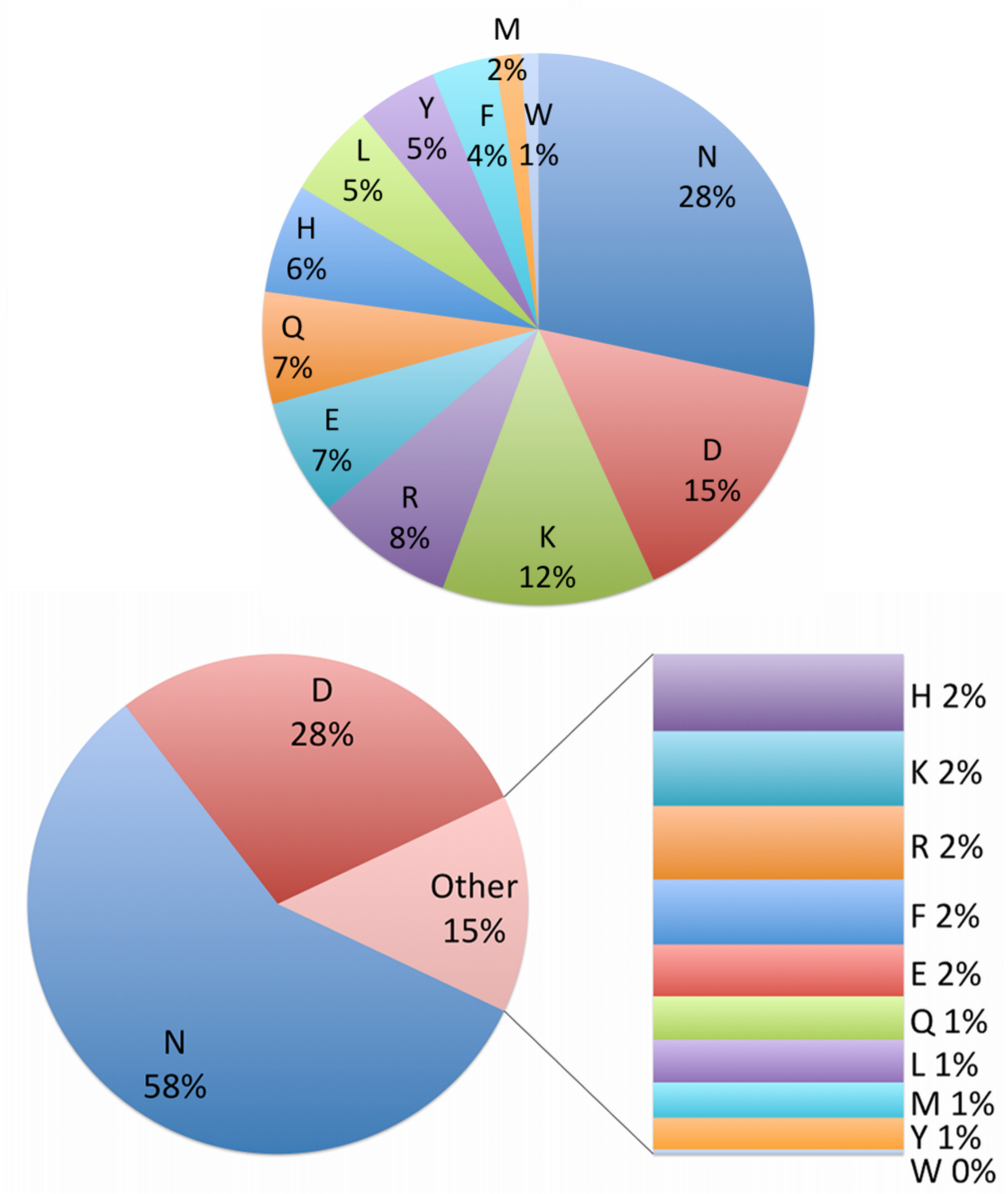}}
    \caption{(Color online:) The relative number of different amino acids in the putative $\it g-$ (top) and $\it trans$ (bottom) $C_\gamma$-islands.}
    \label{fig:figure-6}
  \end{center}
\end{figure}


We proceed to the next level along the side-chain, to map the $C_\delta$ carbons.  in Figure 7 we  plot these 
carbons for those side-chains where $C_\beta$ is located in 
the $\tt L$-$\alpha$ island. In the Figure 7 on top, we show them as they are seen by by our discrete 
Frenet frame observer who sits at the locations of the $C_\alpha$ carbons. In the Figure 7 on bottom we 
show them as they are seen in the C$_\beta$ frame for an observer now sitting at the C$_\beta$ location,
this time using the stereographic projection.
Since  ASN (and ASP)  has no
$C_\delta$ carbon, we display instead the direction of the
side-chain $O$ atom for ASN,  the result is shown in Figure 8.  In the top Figure 8 we use the C$_\alpha$ based 
Frenet frame observer and in the middle and bottom Figure 8 we use the C$_\beta$ frame observer  in combinations
with stereographic projection.


\begin{figure}[!hbtp]
  \begin{center}
    \resizebox{8cm}{!}{\includegraphics[]{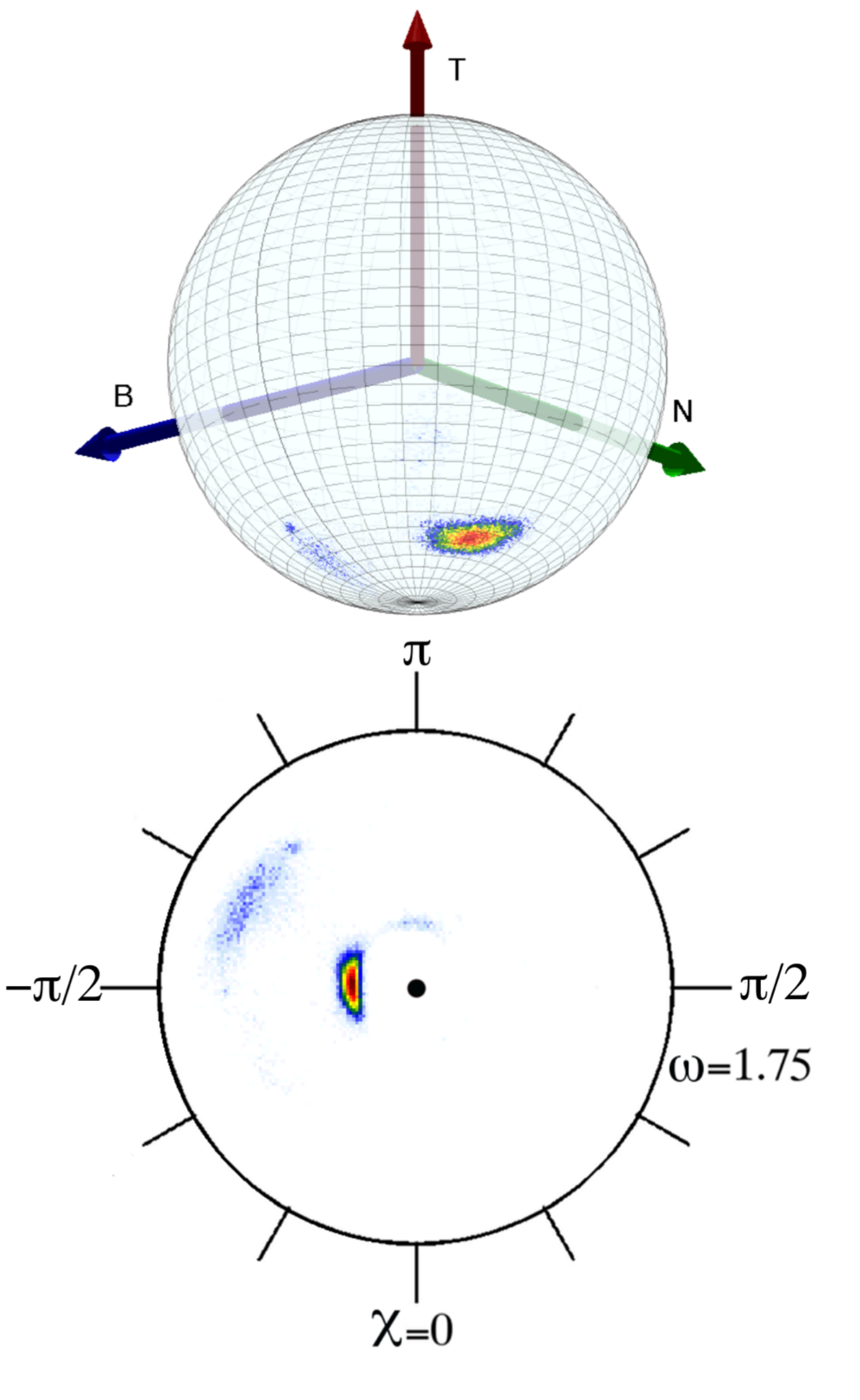}}
    \caption{(Color online:) The directions of the  $C_\delta$-carbons in the discrete Frenet frame of the  $C_\alpha$  carbons on the surrounding 
    two-sphere (top) 
    and in the C$_\beta$ frame (bottom).  In the Figure on top, the C$_\beta$ atom 
    is located at the origin of the  two-sphere, which is stereographically projected
    from the north pole (with C$_\alpha$ at the south pole). } 
    \label{fig:figure-7}
  \end{center}
\end{figure}


From Figure 7 we observe that the directions of the $C_\delta$ continue to be highly localized, independently of the type of
amino acid. But unlike in the case of $C_\gamma$, we find quite surprisingly,  that now there is only one clearly visible  island.
We do not have any definite stereochemistry or  physics based 
explanation why the clearly visible sp3 hybridization based doubling that we observe at the level of $C_\gamma$  has now completely disappeared. However,  
we do observe the formation of a second, relatively very
weakly occupied island at larger values of the latitude angle and with longitude
angle $\chi \sim -2\pi/3$. This island is clearly visible  in Figure 7 (right). There is also a third, very faint island
in the direction $\chi \sim \pi$ that (barely) becomes visible in the stereographically projected Figure 7 (left).
At the moment we do not have a basis to conclude whether the extremely low population of the second and third
island is a real effect or only a reflection of problems in the experimental data.
We refer to \cite{errors}, that there are presently an estimated half a million incorrectly positioned  
side-chain atoms PDB data. In this light, the reason for the sparse population of the two additional
sp3 hybridized  islands should  be subjected to experimental curiosity,  to determine the cause. 


\begin{figure}[!hbtp]
  \begin{center}
    \resizebox{8cm}{!}{\includegraphics[]{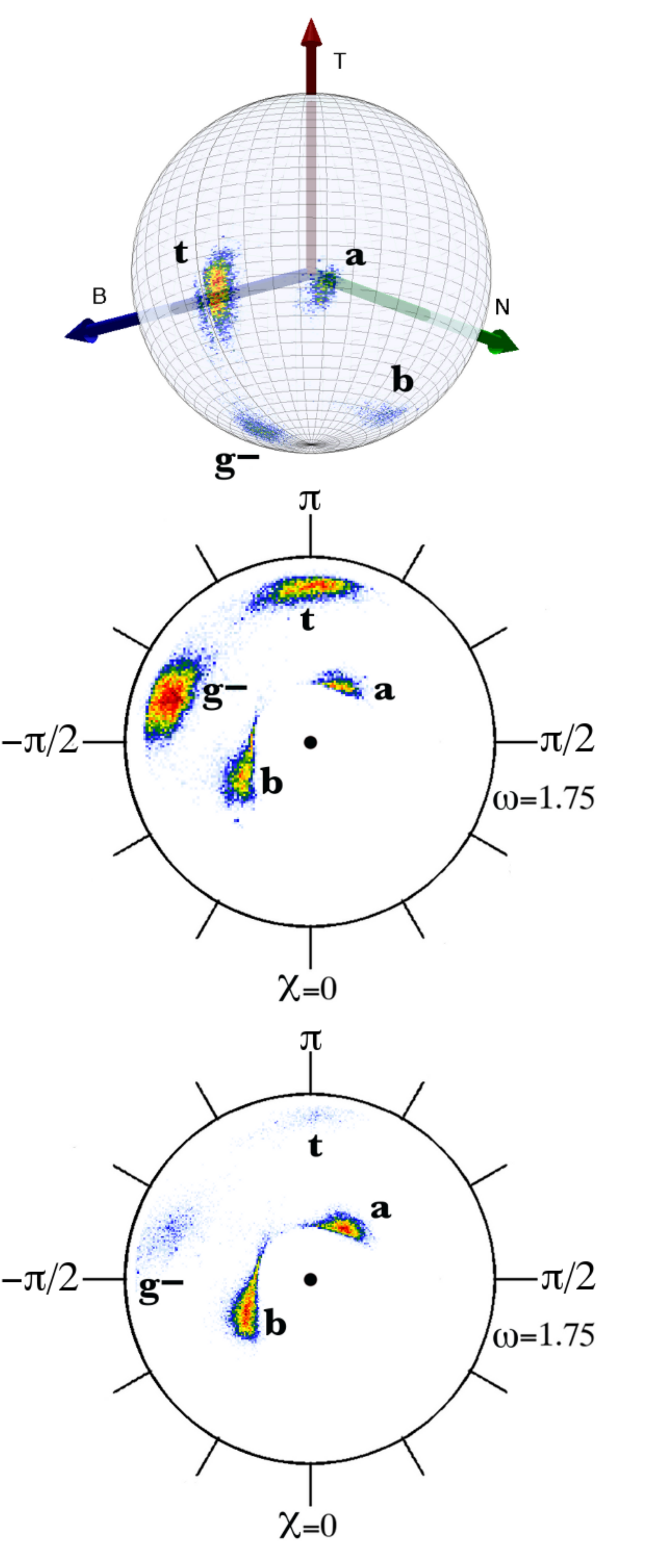}}
    \caption{(Color online:) 
On top, the directions of the  side-chain $O$ atoms of ASN  in the discrete Frenet 
    frame of the  $C_\alpha$  carbons.  In the middle and bottom we use the same stereographically projected C$_\beta$
    frames as in Figure 7 (right). The Figure in middle displays the same atoms as the Figure on top. On bottom, 
    the directions of the  side-chain $N$ atoms of ASN suggests that the correct identification 
    of {\bf a} and {\bf b} regions in the top and middle figure should be  $N$ and not $O$ as in PDB.  Similarly, the regions
    {\bf t} and {\bf g-} in the bottom should be $O$ and not $N$ in PDB. See \cite{errors}.  ({\bf t} is {\it trans} and {\bf g-} is {\it gauche-})}
    \label{fig:figure-8}
  \end{center}
\end{figure}


Since  ASN has no
$C_\delta$ carbon, in Figure 8 we display instead  the $O$ atoms of the ASN side-chain  according to PDB identification.
In the top Figure 8 we use discrete Frenet  frame of the $C_\alpha$-carbons, and in the middle and bottom we use
the stereographically projected C$_\beta$ frame. We note that the two $C_\gamma$ islands appear to become divided into four distinct but still highly 
localized islands. However, we recall that 
the identification between the ASN side-chain $O$ and $N$ can be very difficult,  and  there are apparently 
numerous errors in the $O$ and $N$ identifications in PDB data \cite{errors}. 
Thus we have displayed in Figure 8 (top) the $N$ atoms  according to PDB identification
as well. By comparing the Figures 8 (middle) and (bottom) we propose that {\it most likely}
the two inner-most islands denoted {\bf a} and {\bf b}  in Figure 8 describe $N$ instead of $O$ atoms.
If so, our visualization technique could become a useful tool in detecting erroneously identified $O$ and $N$ atoms
and help to resolve the kind of  issues raised in \cite{errors}. 
This could  be scrutinized by a careful re-analysis of  high resolution 
x-ray crystallography data. 

Finally, in Figure 9 we have mapped the locations of the C$_\gamma$ atoms in our entire
dataset (except for prolines) as they are seen in the stereo graphically projected C$-\beta$  
frame. In the Figure at top we show all atoms  except  those that have C$_\beta$ in the $\tt L$-$\alpha$, 
and in Figure at bottom we show only the $\tt L$-$\alpha$ atoms. The  sp3 hybridization of 
the C$_\beta$ covalent bond structure is clearly visible. Furthermore, in each of the three regions in left we 
recognize the substructure that correspond to the $\alpha$-helices, $\beta$-strands and the interconnecting loops. 
Each of the three regions is then isomorphic to Figure 2a. 

Obviously, it is straightforward to continue the present analysis to inspect additional side-chain atoms. However,
here our goal is not to perform a detailed and complete analysis of all the side-chain atoms, we  simply aim  to 
describe a method.


\begin{figure}[!hbtp]
  \begin{center}
    \resizebox{6cm}{!}{\includegraphics[]{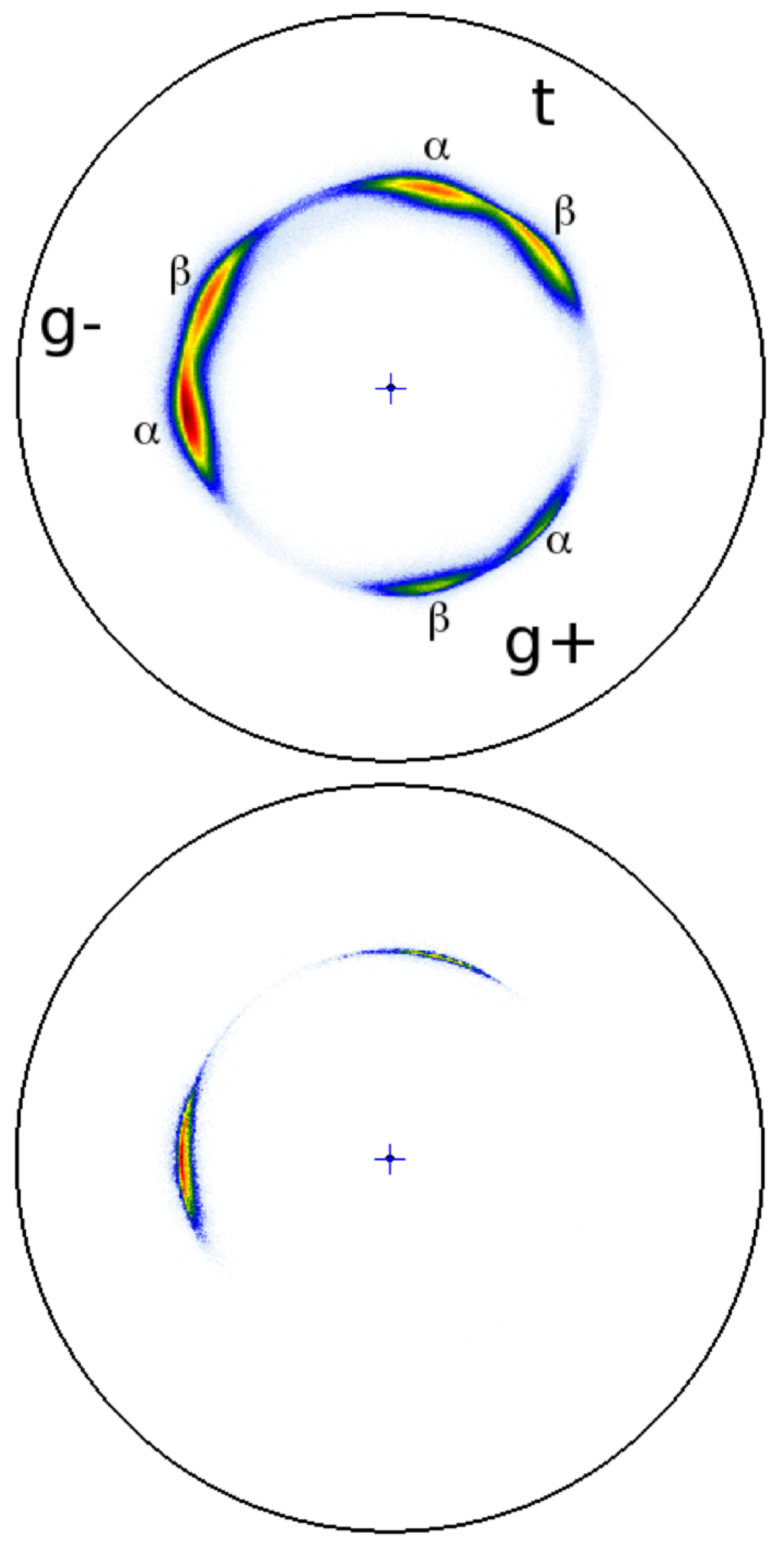}}
    \caption{ (Color online:) The directions of the C$_\gamma$ atoms in the C$_\beta$ centered frames, on two-sphere stereographically
    projected from its north pole and with C$_\alpha$ at south-pole.  
    On the top all those C$_\gamma$ atoms for which the C$_\beta$ is not in
    the $\tt L$-$\alpha$, and on bottom those that correspond to C$_\beta$ in the $\tt L$-$\alpha$ island.  ({\it gauche-} on left,
    {\bf t} on top-right and {\it gauche+} on bottom-right.) 
     }
    \label{fig:figure-9}
  \end{center}
\end{figure}




\section{solitons}


The localization we have observed in the $\tt L$-$\alpha$ 
side-chain atoms proposes that there is an organizational principle in the side-chain orientations
that extends beyond a single peptide unit. Hence it can not be detected by the
Ramachandran plot or in terms of the standard rotamer libraries, these only provide information on
a given peptide unit.  The backbone C$_\alpha$ atoms we have inspected all correspond to the  
$\tt L$-$\alpha$  position of the C$_\beta$, this region is known to commonly appear in 
connection of loops in lieu of regular secondary structures.  Thus the order we have 
observed is {\it a priori} not a reflection of  any apparently regular secondary structure
category at the level of the backbone geometry, but a characteristic of loops.
We  propose that it is due to the presence of a soliton solution to a discrete version of 
nonlinear Schr\"odinger (DNLS) equation that universally describes the backbone C$_\alpha$ 
geometry in (practically) all folded proteins. 

For the soliton description we do not need to know the atomic
level details of the energy function. We only need to apply general symmetry principles to the abstract full quantum mechanical, all-atom Hamiltonian operator $\mathcal H [q_i, p_i]$. 
Here the index i = 1, ..., N labels all pairs of canonical coordinates $(q_i,p_i)$ that 
describe the elementary constituents. These include 
the individual C, O, N, H and every other atom in the protein and in the solvent. We also account for 
the valence electrons, and for every local and long range interaction between all the atoms both in the 
protein and in the solvent. For simplicity we take all the variables to be point-like, that is we work at the
first quantized level. The canonical partition function is computed by the path integral,
\[
\mathcal Z = Tr e^{-\beta \mathcal H } = \int [dq] e^{-\frac{1}{\hbar} \mathcal S(q, \dot q)}
\]
where $\mathcal S(q,\dot q)$ is the classical Euclidean action of $\mathcal H [q_i, p_i]$. The 
integration extends over all period configurations (anti-periodic in the case of fermions).
With the partition function we obtain the thermodynamical Helmholtz free energy as follows: 
We introduce external sources $j_i(t)$ and extend the partition function into the generating functional of 
connected Green's functions,
\[
W[j] = \ln \left \{ \int [dq] e^{ - \frac{1}{\hbar} \mathcal S_\beta + \frac{1}{\hbar} \int\limits_0^\hbar \beta j\dot q } \right\}
\]
We introduce the Legendre transformation
\[
\Gamma[q] = - \frac{1}{\beta} \left \{ W[j] - \int\limits_0^{\hbar \beta} \frac{\delta W}{\delta j}\cdot j \right\}
\]
This defines the effective action that coincides with the Helmholtz free energy 
when we take the limit  $j_i \to 0$. 

There are various methods to compute the Helmholtz free energy $E$. Here we introduce a finite difference
version of the gradient expansion. In the leading nontrivial order we have 
\[
E = \lim\limits_{j_i \to 0} \Gamma[q] = \sum_{i} V^{(i)}(q_i) + F^{(i)} (q_i, q_{i+1}) + \dots
\]
\begin{equation}
= \sum_i V^{(i)}(q_i) + \frac{\partial F^{(i)} (q_i, q_{i+1})}{ {\partial q_{i+1}}_{| q_{i+1}=0}   }  q_{i+1} \ + \dots
\label{grad}
\end{equation}
The potentials $V^{(i)}$ are all local and the  $F^{(i)}  $ are bi-local. The higher order terms
in the expansion are either higher order  polynomials in the nearest neighbor variables,  or terms that
introduce couplings between next-to-nearest neighbor variables.

In the expansion (\ref{grad}), in the case of the backbone 
we identify the generalized coordinates $q_i$ with the bond and torsion 
angles ($\kappa_i, \tau_i$)  in (\ref{kappa}), (\ref{tau}). We assume that all the additional variables 
that appear in the full Hamiltonian operator $\mathcal H [q_i, p_i]$ have been  
integrated over in constructing the partition function. They affect 
the detailed functional form of the coefficients  $V^{(i)}$, $F^{(i)}  $ {\it etc.}  in (\ref{grad}).  
Since (\ref{back}) contains only the $\mathbf t_i$, it is clear that the 
expansion (\ref{grad}) in terms of ($\kappa_i, \tau_i$) must remain invariant if we introduce
a local frame rotation in the normal plane spanned by ($\mathbf n_i , \mathbf b_i$). This 
introduces a strong constraint to its functional form.  In \cite{oma1}, \cite{ulf} this 
has been utilized to show that in the leading order the expansion (\ref{grad}) is uniquely determined. 
It can only contain the following terms \cite{cherno}-\cite{peng}, \cite{ulf},
\[
E = - \sum\limits_{i=1}^{N-1}  2\, \kappa_{i+1} \kappa_i  + \sum\limits_{i=1}^N
\left\{ 2 \kappa_i^2 + q\cdot (\kappa_i^2 - m^2)^2 
\right.
\]
\begin{equation}
\left. 
+ \frac{d_\tau}{2} \, \kappa_i^2 \tau_i^2  -  b_\tau \kappa_i^2  \tau_i - a_\tau  \tau_i   +  \frac{c_\tau}{2}  \tau^2_i 
\right\} 
\label{E1}
\end{equation}
Here the first sum together with the 
three first terms in the second sum coincide with the integrable 
energy function of  the conventional DNLS equation with a potential that displays spontaneous 
symmetry breaking.   The fourth ($b_\tau$) and the fifth ($a_\tau$) terms are the 
{\it only} two lower order nontrivial conserved quantities that appear in the integrable  DNLS hierarchy 
prior to the energy. These are  the momentum and the helicity, respectively.  
The last ($c_\tau$) term is the standard Proca mass term.
The parameters are all global and  specific  only to a super-secondary structure such as 
helix-loop-helix. In particular they are independent of the detailed 
nature of amino acids.

Unlike a force field in molecular dynamics, the energy function (\ref{E1}) does not describe the 
fine  details of the atomary level interactions such as Coulomb, van der Waals, hydrogen 
bonding {\it etc}.  Instead,  like an effective Landau-Lifschitz  theory it describes the properties of 
a folded protein  backbone in terms of universal physical arguments. Full details and motivation 
of (\ref{E1}) are presented in \cite{cherno}-\cite{peng}, \cite{ulf} 

The remarkable property of (\ref{E1}) is  that the torsion angle $\tau_i$  is only subject to
local interactions, all explicit non-local interactions are carried by the bond angle $\kappa_i$. Furthermore, since
$\tau_i$ appears at  most quadratically we can solve for  it in terms of $\kappa_i$,
\begin{equation}
\tau_i = \frac{a_\tau + b_\tau \kappa_i^2}{c_\tau + d_\tau\kappa_i^2}
\label{taueq}
\end{equation}
When we substitute this into the variational equation of  $\kappa_i$ that follows  from (\ref{E1}), we arrive 
at a generalized version of the DNLS equation. Its  soliton solution has been constructed in \cite{cherno}-\cite{peng}.
In particular, it has been observed that the soliton can be approximated by the
discretized version of the soliton solution of the continuum dark NLSE soliton  \cite{davy}-\cite{fadd},
\begin{equation}
\kappa_i  =  \frac{ 
m_{1} \cdot e^{ c_{1} ( i-s)  } - m_{2}  \cdot e^{ - c_{2} ( i-s)}  }
{e^{ c_{1} ( i-s) } +  e^{ - c_{2} ( i-s)}   }
\label{bonds}
\end{equation}
Here the various parameters each have a natural interpretations, see \cite{cherno}-\cite{peng} for a detailed
description: The parameter $s$ determines the backbone site location of the center of the fundamental loop 
that is described by the soliton.  The values of the parameters $m_{1,2} \in [0,\pi] \ mod(2\pi)$ are 
entirely determined by the bond angles of the adjacent helices and strands. 
Finally, {\it only}  the  $c_1$ and  $c_2$ are 
intrinsically loop specific parameters, they specify the length of the loop.
The soliton profile of $\kappa_i$ determines the torsion angles $\tau_i$ by  (\ref{taueq}).
According to \cite{peng} practically all  
PDB proteins can be constructed as the sum of terms of the form (\ref{bonds}), 
in a modular fashion from a relatively small number of soliton profiles.

Following \cite{mart} we argue that  in the Frenet frames, the angular positions of 
the side-chain atoms can be similarly determined 
in terms of the corresponding $\kappa_i$ values only.  For this we denote by  ($\theta, \phi$) the standard 
spherical latitude and longitude angles of the sphere that surrounds the C$_\alpha$ observer. 
We propose that to leading order in the expansion (\ref{grad}), in these coordinates  each of the side-chain atoms 
has an energy function that has the same functional form as the energy function of the backbone torsion angles. Consequently, for  each side-chain atom we introduce  only the following two leading contributions to the energy 
\begin{equation}
E_\theta  =  \sum\limits_{i=1}^N
\left\{  \frac{d_\theta}{2} \, \kappa_i^2 \theta_i^2  -  b_\theta \kappa_i^2  
\theta_i - a_\theta \theta_i   +  \frac{c_\theta}{2}  \theta^2_i 
\right\}   
\label{E2a}
\end{equation}
\begin{equation}
E_\varphi  =  \sum\limits_{i=1}^N
\left\{  \frac{d_\varphi}{2} \, \kappa_i^2 \varphi_i^2  -  b_\varphi \kappa_i^2  \varphi_i 
- a_\varphi \varphi_i   +  \frac{c_\varphi}{2}  \varphi^2_i 
\right\} 
\label{E2b}
\end{equation}
Note that these contributions have been {\it carefully selected} so that they will {\it not} change 
the functional form of(\ref{E1}). The addition of
(\ref{E2a}), (\ref{E2b}) will only redefine the coefficients in the $\kappa_i$ dependent terms in (\ref{E1}) which
does not lead to any change in the underlying soliton structure. In particular, we can still utilize the
approximative soliton profile (\ref{bonds}). 
In parallel with  (\ref{taueq})  the spherical angles  ($\theta_i, \varphi_i$) for each of the side-chain atoms 
are then dynamically determined  by  the DNLS soliton profile of the
backbone bond angles $\kappa_i$,
\[
\theta_i = \frac{a_\theta + b_\theta \kappa_i^2}{c_\theta + d_\theta \kappa_i^2}
\]
\[
\varphi_i = \frac{a_\varphi + b_\varphi \kappa_i^2}{c_\varphi + d_\varphi \kappa_i^2}
\]
The present visual analysis implies that in the case of a $\tt L$-$\alpha$ residue the numerical 
values of both ($ b_\theta ,  d_\theta$) and ($ b_\varphi ,  d_\varphi$) are vanishingly small for 
the C$_\beta$, C$_\gamma$ and C$_\delta$ carbons, and for
the side-chain $N$ and $O$ atoms in the case of ASN and ASP.
But  both ($a_\theta , c_\theta$) and ($a_\varphi , c_\varphi$) have amino acid 
independent, finite and {\it universal} values that can be directly inferred from the Figure 2 (middle), 4, 7 and 8 
respectively,
\[
<\theta_i>_{{\tt L}-\alpha}  \ \approx \frac{a_\theta }{c_\theta }
\]
\[
<\varphi_i>_{{\tt L}-\alpha}  \ \approx \frac{a_\varphi }{c_\varphi }
\]
We now proceed to argue that these universal values can be understood in terms of 
relatively few DNLS solitons. We then show how these solitons can be classified using our 
graphical tools.




\section{soliton visualization}


It  has been argued in the literature  that  in the case of ASN and ASP the $\tt L$-$\alpha$ 
Ramachandran region  become  stabilized by a local but non-covalent attractive interaction 
between the side-chain and backbone carbonyls,  with the backbone oxygen atom in
a special r\^ole \cite{deane}, \cite{review}.  Unlike the Ramachandran plot, our Frenet framing
can provide information on the neighboring peptide units and we have  investigated the directions of 
all  backbone $O$ atoms in our data set,  in a group of residues around the $i^{th}$ side-chain 
$C_\beta$ that  is located in the $\tt L$-$\alpha$ island.   The result shown in Figure 10 displays 
how these $O$ atoms are seen by our Frenet frame observer who is located at the $i^{th}$   
central $C_\alpha$ carbon. Our observer  finds that  the directions of the nearby backbone 
$O$ atoms are very srongly localized and correlated. The localization is residue {\it independent }  
and extends itself over at least four different residues: 

$\bullet$ For the $i-2$ site there is strong localization
with a three-fold degeneracy that is reminiscent of the {\it trans/gauche} (sp3 hybridization) degeneracy.  
The data is consistent with vanishing values of both ($ b_\theta ,  d_\theta$) and ($ b_\varphi ,  d_\varphi$).

$\bullet$ For the site $i-1$ we have very strong localization along the longitudinal ($\varphi_{i-1}$) 
direction, with a tiny oscillation in the latitudinal ($\theta_{i-1}$) direction. For the corresponding energy,
($ b_\varphi ,  d_\varphi$) are again vanishingly small while ($ b_\theta ,  d_\theta$) are now small but non-vanishing.

$\bullet$ For the site $i$ we have a single localized  oscillator in the longitudinal direction. Thus ($ b_\varphi ,  d_\varphi$)
are now small but non-vanishing while  ($ b_\theta ,  d_\theta$) vanish. 

$\bullet$ For the site $i+1$ we again find the three-fold {\it trans/gauche}  
degeneracy:  There are three oscillators
in the longitudinal direction, and they are all located very close to the north-pole. Consequently  ($ b_\theta ,  d_\theta$)
vanish while ($ b_\varphi ,  d_\varphi$) do not. In fact, the $\varphi_{i+1}$ amplitudes are quite large.

The localization pattern of the backbone $O$ atoms 
means that for our Frenet frame observer the backbone geometry around a $\tt L$-$\alpha$ residue
shows very little variations.
Only a very limited set of extended backbone geometries are accessible.   Since  the regime that covers the 
sites from the $(i-2)^{th}$ to  the $(i+1)^{th}$ involves  four sets of bond and torsion angles, each of them 
defined in terms of three {\it resp.} four residues we conclude that the  geometries reflect the 
non-local collective interplay of {\it at least up to seven}  different residue sites along the backbone.  
This is in line with our proposal that the positions of the side-chain atoms  
are determined dynamically by the backbone,  in terms of a small number of different DNLS soliton profiles 
according to (\ref{E2a}), (\ref{E2b}).


\begin{figure}[!hbtp]
  \begin{center}
    \resizebox{8cm}{!}{\includegraphics[]{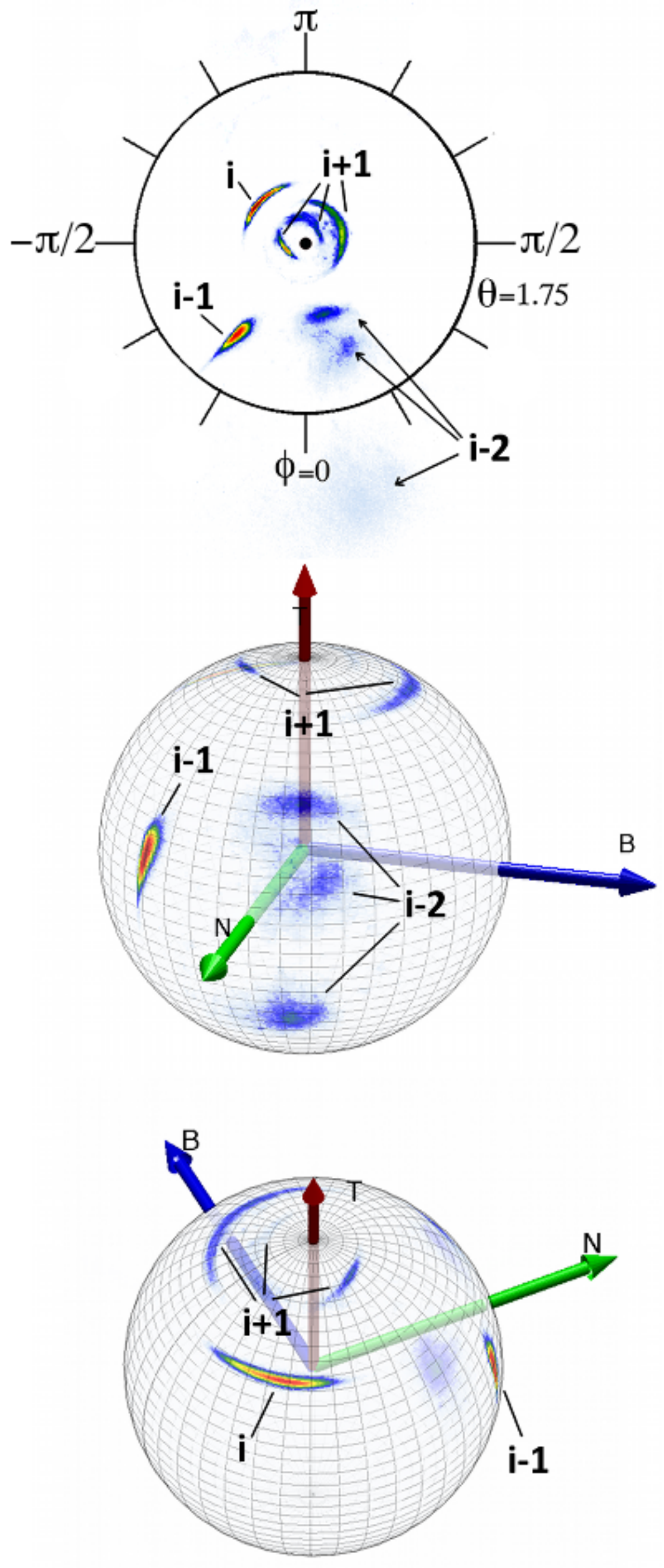}}
    \caption{ (Color online:) The orientations of backbone $O$ atoms around the site $i$ that is located in the  $\tt L$-$\alpha$ island, as seen by our
    discrete Frenet frame observer at the $i^{th}$ central $C_\alpha$-carbon, on a stereographically projected two-sphere (top) and
    on the surrounding two-sphere that we have displayed from two different 
    perspectives  (middle, bottom). For the  $i^{th}$ and $(i-1)^{th}$ atom only one position
    appears to be available while the $(i+1)^{th}$ and $(1-2)^{th}$ atoms each have three available  ({\it trans/gauche}) positions.  
    The angle $\phi$ is measured from the $\bf N$
    axis.  }
    \label{fig:figure-10}
  \end{center}
\end{figure}


To expose the soliton structures  that surround the $\tt L$-$\alpha$ island,  we consider the distribution 
of the backbone bond and torsion angles  that are attached to those $C_\alpha$ carbons where the
$C_\beta$ atom is in the $\tt L$-$\alpha$ position. The result is shown  in Figure 11 on a stereographically projected 
two-sphere, separately for ASN and ASP and for the remaining
 non-glycyl  amino acids.


\begin{figure}[!hbtp]
  \begin{center}
    \resizebox{8.cm}{!}{\includegraphics[]{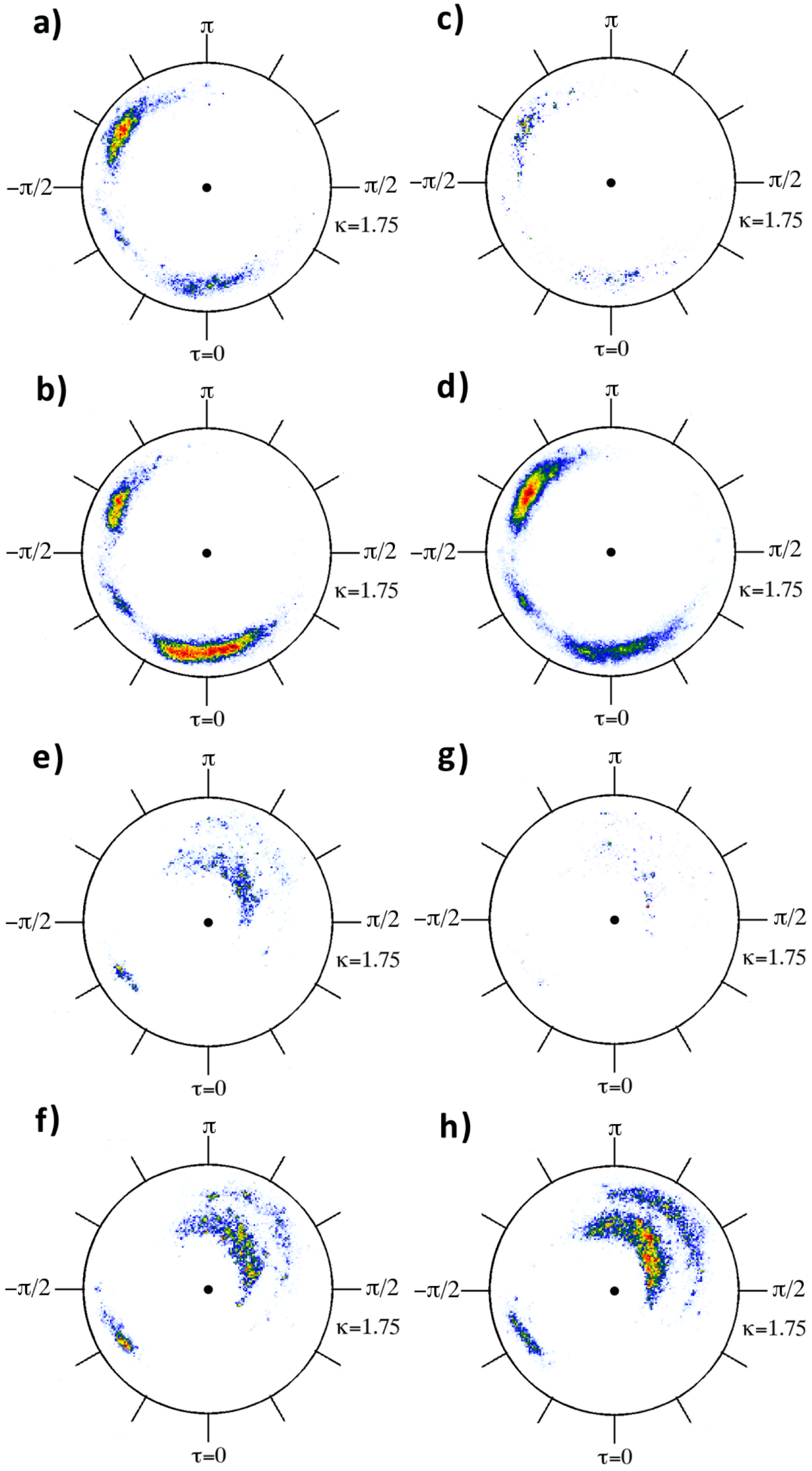}}
    \caption{(Color online:) The $(\kappa,\tau)$ distributions for backbone links that are attached to a $C_\alpha$ carbon with $C_\beta$ in the $\tt L$-$\alpha$ island
    on stereographically projected two-sphere as in Figure 1. We display 
    separately ASN, ASP, and all  the rest. On left column  the $C_\alpha$ carbons 
    in the case where the corresponding $C_\gamma$ carbon is in the  {\it trans} island,
    on right for those where the $C_\gamma$ is in the $g$-  island. First row a),c) is for link that precedes 
    either ASN or ASP. Second row b),d) is
    for link preceding  any other  non-glycyl  amino acid.    Third row e),g) is for link following either ASN or ASP.  Fourth row
    f),h) is for 
    link following the others. }
    \label{fig:figure-11}
  \end{center}
\end{figure}


We observe no practical difference between  the residues. Nor do we find any practical
difference between the various {\it trans} and $g$- positions. 
Instead, we  do observe the following general pattern: 
For  the backbone $C_\alpha$-$C_\alpha$ link that
precedes the $\tt L$-$\alpha$ island,  three different  regions on the 
$(\kappa, \tau)$ plane are probable. These are the regions
that we have denoted with {\bf a}, {\bf b} and {\bf c} respectively  in  the 
Figure 12 (top); In this Figure we have combined all the  
data that are displayed separately in the parts {\bf a, b, c} and {\bf d} of Figure
11.  After the  $\tt L$-$\alpha$ island  there are also three different regions that 
are probable. We denote  these regions  
with letters {\bf b} and {\bf d} and {\bf e} respectively in the Figure 12 (bottom),  
now combining  the data in parts {\bf e, f, g, h} in Figure 11.  
Note that the regions {\bf b} in the two parts of Figure 12 practically overlap. 


\begin{figure}[!hbtp]
  \begin{center}
    \resizebox{8cm}{!}{\includegraphics[]{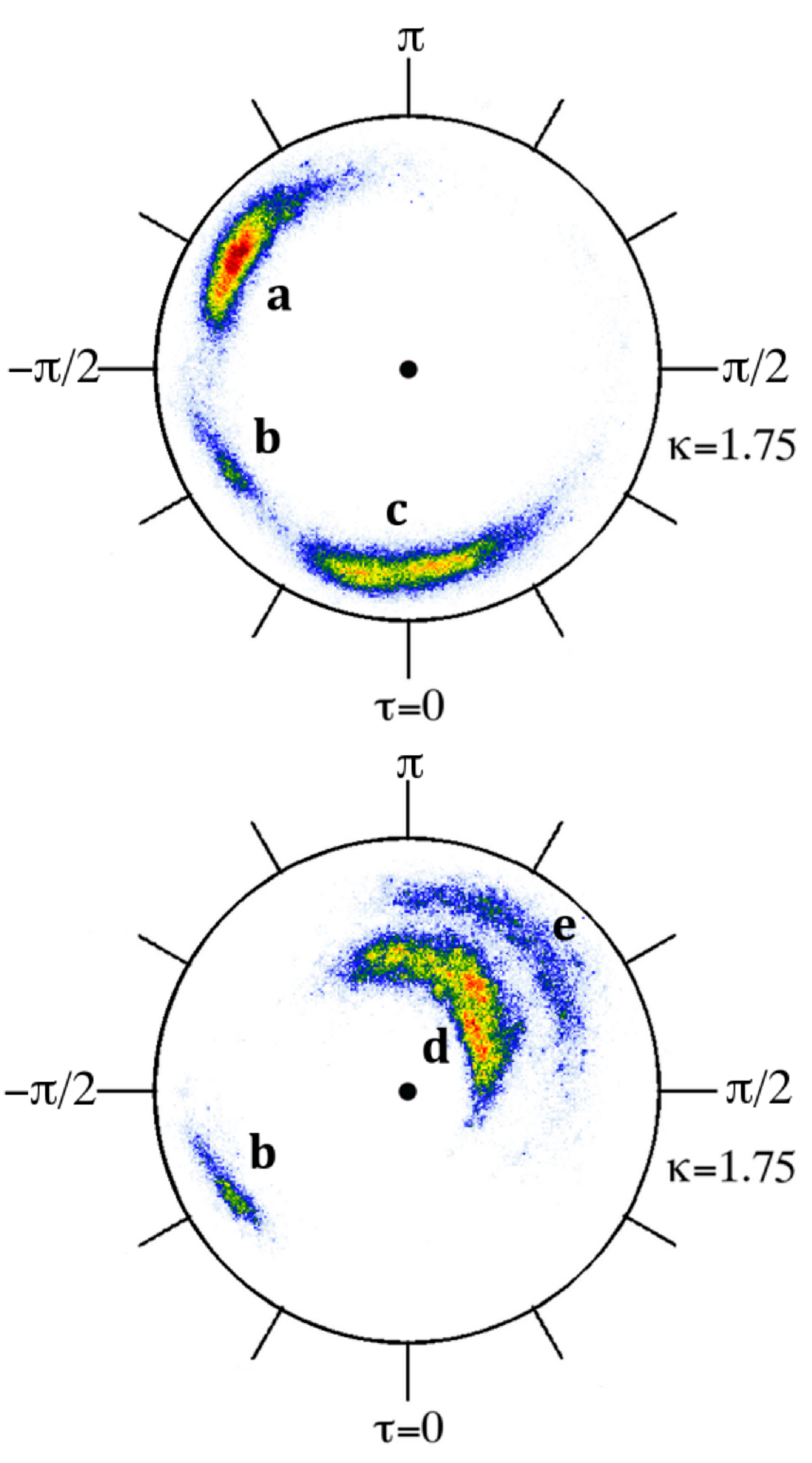}}
    \caption{(Color online:) The $(\kappa,\tau)$ distributions for all backbone links that are attached to a residue in the $\tt L$-$\alpha$ island. On the top for the link preceding
    the residue in the $\tt L$-$\alpha$ island, on the bottom following the residue in the $\tt L$-$\alpha$ island.}
    \label{fig:figure-12}
  \end{center}
\end{figure}


By inspecting the protein structures in our data set we conclude 
that the presence of a residue in the $\tt L$-$\alpha$ island causes the following phenomenological
 {\it selection rules} between the regions displayed in Figure 12. When we roller-coast along the backbone:

\vskip 0.4cm

$\bullet$ The region {\bf a}   can only precede regions {\bf d} and {\bf e}.  

$\bullet$ Both regions {\bf b} and {\bf c} can be followed by any of the three regions {\bf b}, {\bf d} and {\bf e}.

$\bullet$ The  residue preceding either {\bf a} or {\bf c}  is not
located in the $\tt L$-$\alpha$ island. 

$\bullet$ Both the residue preceding  and following {\bf b} can be  located in the  $\tt L$-$\alpha$ island. 

$\bullet$ If the two residues following {\bf c}  are both  in the  $\tt L$-$\alpha$ island, the 
first  residue  connects {\bf c} 
to {\bf b} and the second connects  {\bf b} to {\bf b}. 
 
\vskip 0.4cm

These are the selection  rules, that limit 
the available global topology of the backbone solitons when we pass the $\tt L$-$\alpha$
site.  Notice that since the  region {\bf b} has the same bond angle as the standard $\alpha$-helix region and
since the torsion angles are  equal in magnitude but have an opposite  sign, a repeated structure in {\bf b} is the
right-handed mirror image of the standard $\alpha$-helix. Consequently this is truly the region of  {\it left-handed $\alpha$-helices}.

These selection rules classify the different possible DNLS soliton profiles in
the presence of the $\tt L$-$\alpha$ island.  In particular, we have
found that there seems to be no more than four solitons that are particularly common around  a residue with 
$C_\beta$ located in the  $\tt L$-$\alpha$ island.  We now describe these solitons qualitatively using our visual tools,  as transition trajectories 
between the different regions that appear in the maps of Figure 1. These transitions 
illustrate  how our observer turns at the location of each $C_\alpha$ as  she roller-coasts
through the soliton. The results are shown in  Figure 13. In each case the pink arrow corresponds to a site where $C_\beta$ 
is located in the $\tt L$-$\alpha$  
island; Recall that the bond and torsion angles are link variables, 
they connect two  $C_\alpha$ carbons according to (\ref{DFE}). Clearly, the Figure 13 is but an example of a general 
method to visually classify solitons in a manner that directly reflects the geometry of the underlying backbone. 


\begin{figure}[!hbtp]
  \begin{center}
    \resizebox{4.5cm}{!}{\includegraphics[]{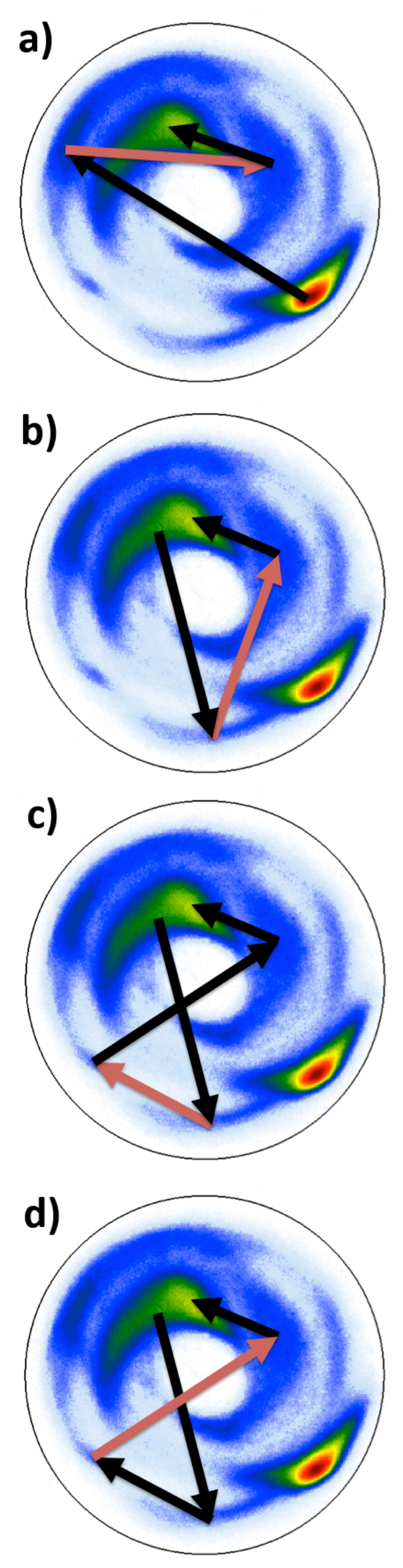}}
    \caption{(Color online:)  Four different soliton trajectories through a residue in the $\tt L$-$\alpha$ island that are common in our data set, on the stereographically projected two-sphere. The arrow shows how the Frenet frame observed sees the soliton to 
    proceed from a $C_\alpha$ to the 
    next $C_\alpha$. In each case the pink line denotes 
    the transition that is caused by  the presence of a residue in the $\tt L$-$\alpha$  island. The trajectory  {\bf a} described a soliton that connects
    the $\alpha$-helix region to the $\beta$-strand region.    
    The remaining ones all both start and end in the   $\beta$-strand  region.  }
    \label{fig:figure-13}
  \end{center}
\end{figure}


$\bullet$ In Figure 13a, the first residue takes our observer  away from an $\alpha$-helix region  to the region
{\bf a} in the Figure 12 left (black arrow).  This is followed by a residue in the $\tt L$-$\alpha$ island, that takes the observer 
to the region {\bf d} in
the Figure  12 right (pink arrow). Finally, there is  a transition to the $\beta$-strand region (black arrow).  Consequently this is a short  soliton that
takes us from the ground state which is an $\alpha$-helix to the other ground state which is a $\beta$-strand. 

$\bullet$ The second soliton  trajectory shown in Figure 13b starts from the $\beta$-strand region with a residue that  takes the observer into region {\bf c} in Figure 12 left. 
The following  residue  that is located in the $\tt L$-$\alpha$ island then causes a transition into region {\bf d} in Figure 12 right (pink arrow).
This is followed by a transition back to a $\beta$-strand region. Since the initial and final positions are in a $\beta$-strand, 
this is an example of a soliton that combines the $\beta$-strand with another $\beta$-strand.

$\bullet$ The third trajectory that we have described in Figure 13c starts from the $\beta$-strand region and proceeds to region {\bf  c} in Figure
12 left.  From there  the trajectory proceeds to region {\bf b}
in Figure 12 left,  with the transition caused by a residue in the $\tt L$-$\alpha$ island. This is followed by a transition to
region {\bf d} and then back to the $\beta$-strand region.  This trajectory is also an example of a  soliton that combines 
the $\beta$-strand with another $\beta$-strand.

$\bullet$ Finally, the fourth trajectory that is also common in our data set is the one displayed in Figure 13d. It is similar with the trajectory
described in Figure 13c, except that now the residue that is located in $\tt L$-$\alpha$ island causes the transition from {\bf b} to {\bf d}
in Figure 12. This trajectory is also an example of a  soliton that combines the $\beta$-strand with another $\beta$-strand.

The remarkable property of solitons c) and d) in Figure 13 is, that they have similar overall topology and differ from each other only by the location of the $\tt L$-$\alpha$ along the trajectory. It is quite plausible that in some proteins 
these two solitons are but two states of an oscillating  discrete "breather" soliton. The ensuing
proteins are  presumable unstructured.

Finally, for the purposes of soliton taxonomy we note that 
when we analyze the proteins in the version v3.3 Library of chopped PDB files for representative  
CATH domains we find that 
the  propensity of our solitons is largest  in the (mainly-$\beta$) CA level classes 2.90, 2.160  where over 5$\%$ of all  residues are in the 
 $\tt L$-$\alpha$ island. We also find that  any CA level family has  at least 1$\%$ of their residues in the island, except 1.40 where the 
single representative with PDB code 1PPR  has no residues in the island.

\section{Conclusion}

We have developed a new visualization method of proteins. In the case of backbones our method provides information
about the geometry of neighboring peptide units. This enables us to go beyond the regime of  the canonical
Ramachandran plot which does not contain information on the neighboring units. 
As an example of side chains, we have visually investigated the non-glycyl  residues that are located 
in the $\tt L$-$\alpha$ region of the Ramachandran plot. {\it Independently} of the amino acid,  we find that for a discrete Frenet frame observer who roller-coasts along the backbone 
$C_\alpha$ carbons  the corresponding side-chain  $C_\beta$ carbons are always localized  
in the same direction which is clearly different from the direction of the $C_\beta$ carbons in the right-handed region.  
This universality in the orientation persists  when we  investigate the $C_\gamma$ and $C_\delta$ carbons, and
the side chain $O$ and $N$ atom  in the case of ASN and ASP. 
The results  suggest that instead of reflecting {\it only} a local interaction between
a given backbone unit and its residue,  the $\tt L$-$\alpha$ island  
is also associated with  a largely residue independent  backbone conformation.

 When we proceed to analyze the distribution of those backbone bond and torsion angles  that are associated with
the links that both precede and follow a residue that is located in the   $\tt L$-$\alpha$ island, we find that  independently of the residue
these angles display very similar patterns. Since the definition of a  bond angle takes three $C_\alpha$ carbons 
and the definition of a torsion angle takes four, this prompts us to propose that the geometrical structure
associated with the presence of a residue in the  $\tt L$-$\alpha$ island  is a  soliton that reflects the 
interplay of at least seven  consecutive backbone units. In particular, we have not been able to pin-point any  
obvious local  reason (charged, polar, acidic, hydrophobic/philic) to explain the presence or absence of a residue   
on the $\tt L$-$\alpha$ region. 

Our approach is based on a novel  visualization method to depict proteins. This method is based on advances
in three dimensional visualization techniques that have been 
developed after Ramachandran presented his plot.  In the course of our analysis we have been able to
observe several systematic patterns including potential anomalies in the PDB data. The visualization
method we have developed shows promise to
become a valuable tool for both experimental and theoretical protein structure analysis  and fold description, in particular
for visually describing and classifying the backbone solitons and as a complement to existing side-chain 
rotamer libraries.
 
\vskip 0.5cm 
\section*{Acknowledgement}
We thank S. Hu and J. \.Aqvist for discussions.


\vskip 0.5cm


\begin{thebibliography}{}

\bibitem{rama1} G. Ramachandran, C. Ramakrishnan and V. Sasisekharan,  
Journ.  Mol. Biol.  {\bf  7}  95
(1963)

\bibitem{rama2} C. Ramakrishnan and G. Ramachandran, 
Biophys. J. {\bf 5} 909 (1965)

\bibitem{janin} J. Janin, S. Wodak, M. Levitt, D. Maigret, J Mol Biol.  {\bf 125} 357 (1978)


\bibitem{hansonbook} A.J. Hanson,     {\it Visualizing Quaternions}, Morgan Kaufmann Elsevier (London, 2006)

\bibitem{kuipers} J.B. Kuipers,   {\it Quaternions and Rotation Sequences: a Primer with Applications to Orbits, Aerospace, and Virtual Reality}, Princeton University Press (Princeton, 1999) 

\bibitem{cherno}  M.N.Chernodub, S. Hu and  A.J.   Niemi 
Phys.  Rev.  {\bf E82 } 011916 (2010)


\bibitem{nora}  N.Molkenthin, S. Hu and A.J. Niemi 
Phys. Rev. Lett. {\bf  106}  078102 (2011)

\bibitem{peng} A. Krokhotin, A.J. Niemi and X. Peng, arXiv:1109.3903v1 [physics.bio-ph]

\bibitem{davy} A.S.  Davydov, 
Journ.  Theor.  Biol. {\bf 66} 379 (1977)

\bibitem{kevk} P.G. Kevrekidis,  {\it The Discrete Nonlinear Schr�dinger Equation: Mathematical Analysis, Numerical Computations and 
Physical Perspectives} (Springer-Verlag, Berlin, 2009) 


\bibitem{fadd} L.D. Faddeev and L.A. Takhtajan, {\it Hamiltonian methods in the theory of solitons}  (Springer
Verlag, Berlin, 1987)


\bibitem{dff} S.Hu, M. Lundgren and A.J.  Niemi 
Phys. Rev. {\bf E83} 061908 (2011)



\bibitem{deane} C.M. Deane, F.H. Allen, R. Taylor and T.L.  Blundell, 
Prot. Eng. {\bf 12} 1025  (1999)

\bibitem{allen} F.H. Allen,  C.A. Baalham,  J.O.M. Lommerse  and P.R. Raithby,
Acta Crystallog. {\bf  B54}   320 (1998)

\bibitem{review} P. Chakrabarti and D. Pal,
Prog. Biophys. Molec. Biol. {\bf 76} 1 (2001)


\bibitem{deami}  N.E. Robinson and A.B.  Robinson, 
PNAS (USA) {\bf 98} 12409  (2001)

\bibitem{race}  C.R. McCuddena and V.B. Kraus, 
Clinic.  Biochem.  {\bf 39}   1112 (2006)

\bibitem{deami2}   N.E. Robinson and A.B.  Robinson,  {\it Molecular Clocks � Deamidation of Asparaginyl and 
Glutaminyl Residues in Peptides and Proteins}
Althouse Press (London, 2004)


\bibitem{prion}  E.H. Koo,  P.T. Lansbury and J.W.  Kelly, 
PNAS (USA) {\bf 96} 9989 (1999)



 \bibitem{bishop} R.L. Bishop, 
 Am. Math.  Monthly {\bf 82} 246-251 (1974)


\bibitem{dh} R.S. Hartenberg and J.  Denavit,  {\it Kinematic synthesis of linkages} McGraw-Hill (New York, NY, 1964)

 
 \bibitem{pdb}  H.M. Berman, K.  Henrick, H. Nakamura and J.L.  Markley,  
{\it Nucl. Acids Res. } {\bf 35}  (Database issue) D301 (2007)

\bibitem{cath} C.A. Orengo, A.D.  Michie, S. Jones, D.T. Jones, M.B. Swindells and J.M.
Thornton, 
Structure {\bf 5} 1093 (1997)

\bibitem{sch}  L. Sch\"afer and L.M.   Cao,  
Journ. Mol. Struc. {\bf 333} 201 (1995)	

  
\bibitem{krp1}  P.A. Karplus, 
Prot. Sci. {\bf 5} 1406 (1996)

\bibitem{krp2} D.S. Berkholz, M.V. Shapovalov, R.L.  Dunbrack,  Jr. and P.A. Karplus, 
Structure {\bf 17} 1316 (2009)

\bibitem{tou} W.G. Touw and G. Vriend, 
Acta Cryst.  {\bf D66} 1341 (2010)

\bibitem{mart} M. Lundgren and A.J. Niemi,  arXiv:1109.0423v1 [q-bio.BM]

\bibitem{errors}  C.X. Weichenberger and M.J. Sippl, 
Nucl. Acids Res. {\bf 35} (Web Server Issue) W403 (2007)
 
\bibitem{oma1} A.J. Niemi, Phys. Rev. {\bf D67} 106004 (2003)

\bibitem{ulf}  U.H. Danielsson, M Lundgren and A.J. Niemi,  
Phys.  Rev. {\bf E82}  021910 (2010)


\end{thebibliography}
\end{document}